\documentclass[a4,amssymb,superscriptaddress,9pt,longbibliography,twocolumn]{revtex4-1}
\usepackage[pdftex]{graphicx}
\usepackage[british]{babel}
\usepackage[autostyle]{csquotes}
\usepackage{subfig}
\usepackage[toc,page]{appendix}
\usepackage{amsmath,amsfonts,amssymb,mathtools}
\usepackage{diagbox}
\usepackage{float}
\usepackage{hyperref}
\usepackage{nicematrix}
\usepackage{float}
\usepackage{bbm,bm}
\usepackage[normalem]{ulem}
\usepackage{comment}
\usepackage{physics}
\usepackage{xcolor}
\usepackage[utf8]{inputenc}

\usepackage{pgfplots}

\usepgfplotslibrary{groupplots}
\usepgfplotslibrary{fillbetween}
\usetikzlibrary{patterns}

\pgfplotsset{
    every first x axis line/.style={},
    every first y axis line/.style={},
    every first z axis line/.style={},
    every second x axis line/.style={},
    every second y axis line/.style={},
    every second z axis line/.style={},
    first x axis line style/.style={/pgfplots/every first x axis line/.append style={#1}},
    first y axis line style/.style={/pgfplots/every first y axis line/.append style={#1}},
    first z axis line style/.style={/pgfplots/every first z axis line/.append style={#1}},
    second x axis line style/.style={/pgfplots/every second x axis line/.append style={#1}},
    second y axis line style/.style={/pgfplots/every second y axis line/.append style={#1}},
    second z axis line style/.style={/pgfplots/every second z axis line/.append style={#1}},compat=1.12
}
\definecolor{blueprl}{RGB}{46,48,146}
\usepgfplotslibrary{groupplots}

\usepackage{lipsum}
 \usepackage{textcomp} 
\def\ANU{Centre for Quantum Computation and Communication Technology, Department of Quantum Science, Australian National University, Canberra, ACT 2601, Australia.}
\def\NTU{School of Physical and Mathematical Sciences, Nanyang Technological University, Singapore 639673, Republic of Singapore}

\def\CGA{Centre for Gravitational Astrophysics (CGA), Research School of Physics, The Australian National University, Canberra ACT 2601, Australia}
\def\OzGrav{ARC Centre of Excellence for Gravitational Wave Discovery (OzGrav), Research School of Physics, The Australian National University, Canberra
ACT 2601, Australia}

\usepackage{pgfplots}
\usepgfplotslibrary{groupplots}

\usepackage[T1]{fontenc}

\begin{document}
\title{Enhancing the precision limits of interferometric satellite geodesy missions}

\author{Lorc{\'a}n O. Conlon}
\email{lorcan.conlon@anu.edu.au}
%\author{LC}
\affiliation{\ANU}%
\author{Thibault Michel}
\affiliation{\ANU}
%\affiliation{\LKB}
\author{Giovanni Guccione}
\affiliation{\ANU}
\author{Kirk McKenzie}
\affiliation{\CGA}
\affiliation{\OzGrav}
\author{Syed M. Assad}
\affiliation{\ANU}
\affiliation{\NTU}
\author{Ping Koy Lam}
\affiliation{\ANU}
\affiliation{\NTU}
\date{\today}

\begin{abstract}
Satellite geodesy uses the measurement of the motion of one or more satellites to infer precise information about the Earth\textquotesingle s gravitational field. In this work, we consider the achievable precision limits on such measurements by examining approximate models for the three main noise sources in the measurement process of the current Gravitational Recovery and Climate Experiment (GRACE) Follow-On mission: laser phase noise, accelerometer noise and quantum noise. We show that, through time-delay interferometry, it is possible to remove the laser phase noise from the measurement, allowing for almost three orders of magnitude improvement in the signal-to-noise ratio. Several differential mass satellite formations are presented which can further enhance the signal-to-noise ratio through the removal of accelerometer noise. Finally, techniques from quantum optics have been studied, and found to have great promise for reducing quantum noise in other alternative mission configurations. We model the spectral noise performance using an intuitive 1D model and verify that our proposals have the potential to greatly enhance the performance of near-future satellite geodesy missions.
%\clearpage
\end{abstract}
\maketitle
%%%%%%%%%%%%%%%%%%%%%%%%%%  body  %%%%%%%%%%%%%%%%%%%%%%%%%%
\section{Introduction}

The possibility of using a pair of satellites to measure the Earth's gravitational field was first proposed by Wolff in 1969~\cite{wolff1969direct}. Based on this premise the GRACE mission was launched in 2002, providing scientists with the tools necessary to recover the Earth's gravitational field with unprecedented precision~\cite{wahr2004time, tapley2004grace,tapley2004gravity}. GRACE consisted of two satellites which orbited the Earth on very similar trajectories, with an on-board ranging system which measured the satellite separation to great accuracy. The original GRACE mission used a microwave ranging system~\cite{kim2003simulation}, and the second generation mission, GRACE Follow-On (GRACE-FO), included the addition of a laser ranging interferometer (LRI)~\cite{abich2019orbit, sheard2012intersatellite}. Even though the LRI was not designed to be the main instrument in GRACE-FO, and was included to demonstrate improved sensitivity for future missions, it provided a promising indication of future precision enhancement~\cite{abich2019orbit}. This is the first intersatellite optical interferometer, and also serves as an important technological demonstration for the Laser Interferometer Space Antenna (LISA)~\cite{amaro2017laser}, a planned space-borne gravitational wave detector. 

The advantage of a satellite-based LRI is not limited to metrological missions. There are many reasons to believe the future of the quantum internet lies in space~\cite{khatri2019spooky, aspelmeyer2003long, simon2017towards}, and GRACE-FO with its LRI represents an important step towards this vision. On this front, there has been much progress towards a space-based quantum key distribution network~\cite{vallone2015experimental, liao2017satellite, liao2018satellite, bedington2017progress}, and it is only a matter of time before satellite-to-satellite links are employed to greatly extend the distance for secure communication. The GRACE-FO mission already demonstrates some crucial elements of continuous variable quantum communications; both relying on coherent laser links over large distances. Thus, the mission is of great importance, even beyond its contribution to our knowledge of the Earth's gravitational field. 

The interferometric measurement used on GRACE-FO works by measuring the relative phase, in cycles, between the lasers on-board each satellite. Such a measurement intrinsically has two fundamental noise sources: laser phase noise~\cite{abich2019orbit}, caused by imperfect laser stability, and unavoidable quantum noise~\cite{caves1981quantum} caused by photon number fluctuations. In addition to the LRI, the GRACE-FO mission requires accelerometers on board both satellites to distinguish gravitational (signal) and non-gravitational (noise) accelerations~\cite{christophe2015new}. The non-gravitational accelerations come from a variety of sources, such as aerodynamic drag and solar radiation pressure. It is necessary to remove the non-gravitational accelerations from the measurement in order to get a faithful estimate of the gravitational field, hence non-gravitational accelerations can be thought of as another noise source. This noise can be removed using the accelerometer measurement data at the expense of introducing accelerometer instrument noise. We shall use the term accelerometer noise for any noise associated with the non-gravitational acceleration and its removal, i.e. both accelerometer instrument noise and non-gravitational accelerations. Thus, the total measurement noise comes from the accelerometer noise, as well as the laser phase noise and quantum noise from the interferometric measurement. Although in this paper we only consider measurement noise, there are other noise sources which may limit the gravitational field recovery, such as aliasing noise~\cite{dobslaw2016modeling} and tilt-to-length coupling error~\cite{wegener2020tilt}. %Aliasing noise is not, however, a fundamental problem and can be reduced with an array of satellites. 

This paper is divided into three analyses, discussing the possibility of diminishing the effects of each of the measurement noise sources in turn. First we show that time delay interferometry (TDI), which has been proposed for LISA~\cite{tinto1999cancellation, armstrong1999time, tinto2002time, tinto2003implementation, tinto2004time}, is a powerful tool for mitigating the effects of laser phase noise. TDI has been considered before for GRACE-FO, however not to enhance the GRACE-FO mission but as a technological demonstration for LISA~\cite{francis2015tone}. %We then consider how TDI and common mode non-gravitational acceleration cancellation can be used to reduce accelerometer noise and laser phase noise simultaneously. 
We also show that appropriate formations of different mass satellites can be used to reduce accelerometer noise and laser phase noise simultaneously. Multi-satellite formation flying has been suggested~\cite{sneeuw2005satellite,sharifi2007gravity,reubelt2010quick}, however not as a technique for removing measurement noise but to enhance the gravitational signal. Finally we turn to a quantum-limited GRACE, considering what happens when quantum noise is the dominant noise source of such a mission. In this situation, techniques from quantum optics can reduce the quantum noise and we find a whole new regime for satellite geodesy. Indeed, when quantum noise limited, the optimal satellite separation could shrink from hundreds of kilometres to a few kilometres. This suggests that future gravitational recovery missions, perhaps in other planetary settings, may look very different from today's GRACE-FO mission.
\begin{figure*}[t]
\includegraphics[width=\textwidth]{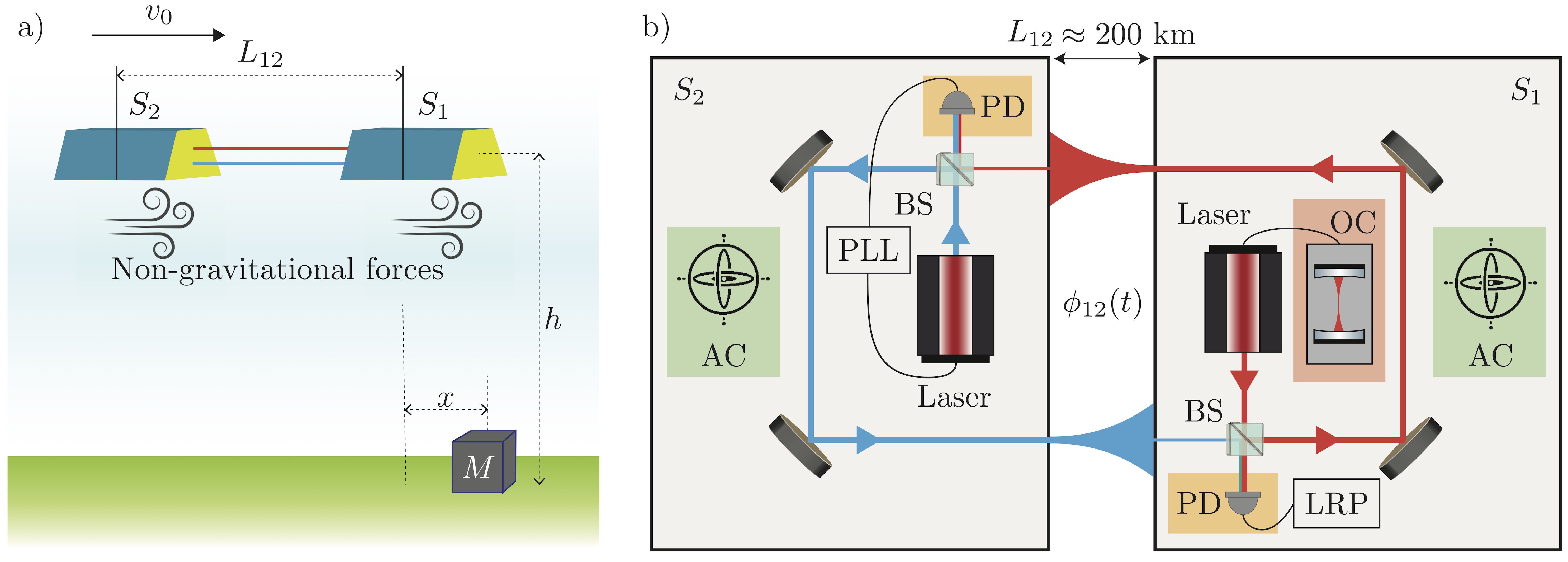}
\caption{\textbf{Schematic of satellite geodesy}.  a) Current GRACE-FO formation. Two satellites separated by a distance $L_{12}$ fly over a mass $M$ at a height $h$ above the Earth. The differential acceleration allows information about the gravitational field to be recovered. b) Detailed schematic showing the main noise sources in the current GRACE-FO mission, highlighted in different colours. Satellite $S_1$ is designated as the master satellite and sends out a laser beam stabilised to an optical cavity (OC), which determines the laser phase noise (red box). The second satellite, $S_2$, returns a laser beam phase locked to this at a 10 MHz offset. Inherent fluctuations in the number of photons manifest as quantum noise, highlighted at the photodetection stage (orange boxes). There are also non-gravitational forces which affect the motion of the two satellites (highlighted in blue in Fig.~a)). Non-gravitational acceleration arises from a variety of sources, including aerodynamic drag and solar radiation pressure. Non-gravitational acceleration can be measured and removed at the expense of introducing accelerometer instrument noise (green box). Measurement instruments include accelerometer (AC), photodetector (PD), beam-splitter (BS), phase locked loop (PLL) and laser ranging processor (LRP).}
\label{fig:scheme}
\end{figure*}

\begin{figure}[t]
\includegraphics[width=.5\textwidth]{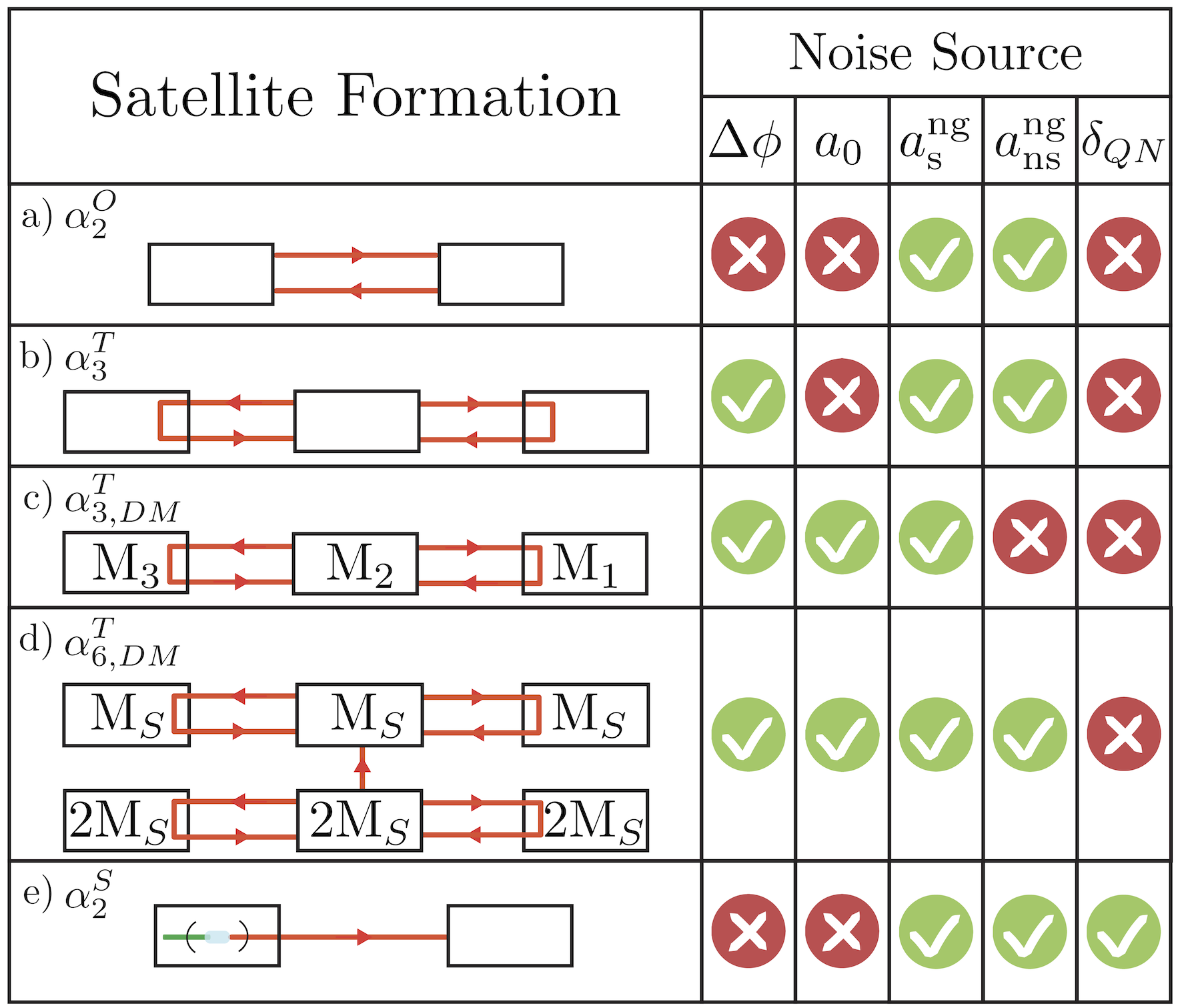}
\caption{\textbf{Possible future mission formations}. Possible formations are denoted by $\alpha^i_{j,k}$, where $i$ represents the technique being used, $j$ represents the number of satellites used and $k$ indicates if satellites of different mass are necessary. a) Original two satellite formation $\alpha^O_2$, b) three satellite formation with TDI $\alpha^T_3$, c) three satellite differential mass formation with TDI $\alpha^T_{3,DM}$, d) six satellite differential mass formation with TDI $\alpha^T_{6,DM}$ and e) two satellite formation using squeezed light $\alpha^S_2$. These allow the removal of various noise sources; laser phase noise $\Delta\phi$, accelerometer instrument noise $a_0$, stationary  and non-stationary non-gravitational accelerations, $a_{\text{s}}^{\text{ng}}$ and $a_{\text{ns}}^{\text{ng}}$ respectively, and quantum noise $\delta_{QN}$. Formations $\alpha^T_{3}$, $\alpha^T_{3,DM}$ and $\alpha^T_{6,DM}$ all use more than two satellites and so can remove laser phase noise through TDI. Formations $\alpha^O_{2}$, $\alpha^T_{3}$ and $\alpha^S_{2}$ use accelerometers and so the $a_{\text{s}}^{\text{ng}}$ and $a_{\text{ns}}^{\text{ng}}$ terms are removed from the measurement at the expense of accelerometer instrument noise. Formations $\alpha^T_{3,DM}$ and $\alpha^T_{6,DM}$ do not use accelerometers and the non-gravitational accelerations are removed through appropriate combinations of the measurements, made possible by the different satellite masses. No scheme can completely remove quantum noise as each additional measurement adds a new source of quantum noise, however it can be reduced through the use of squeezed light as shown in formation $\alpha^S_{2}$.}
\label{fig:multscheme}
\end{figure}

\section{Results}
%Before presenting our main results, we first describe the parameters and noise budget of present satellite geodesy missions.
Before presenting our main results, we first describe the models we shall use for the gravitational signal and measurement noise.
%\subsection{GRACE-FO parameters}
\subsection{Gravitational signal}
GRACE-FO measures sub-micrometer changes in the satellite separation through changes in the phase of the laser light travelling between the satellites. The phase change is then converted to a change in the separation, or range, between the two satellites, which is in turn converted to a range acceleration. The measured non-gravitational accelerations, along with other forces, such as tidal gravitational forces~\cite{savcenko2012eot11a} and other non-tidal forces~\cite{kvas2019grace} which contribute to the background gravitational field, are then removed from this range acceleration. The remaining range acceleration of the two satellites is used to estimate the Earth's local gravitational field. 

In reality this is done considering a spherical harmonic expansion of the Earth's gravitational potential. Instead, we turn to a simpler linear model~\cite{spero2021point} to obtain analytic solutions for the motion of a body in such a field. Although we are primarily concerned with the measurement noise, which is largely unaffected by this simplification, this simplified model may fail to capture the full complexity of real-world satellite geodesy and instead provides an indication of what techniques may be beneficial in reality. A schematic of this model is shown in Fig.~\ref{fig:scheme} a). We consider two satellites at a height $h$ above the ground, separated by a distance $L_{12}$. The first satellite is a distance $\sqrt{h^2+x^2}$ from a point mass $M$ located on the surface of the Earth. Both satellites are initially travelling with velocity $v_0$. In the frequency domain the measured range acceleration between the two satellites is given by Ref.~\cite{spero2021point} as
 \begin{equation}
 \label{eq:acc_twosat}
\abs{a_{\text{R}}(f)}=\frac{16\pi f G\text{M}}{v_0^2}\abs{K_0\left( \frac{2\pi f}{f_h} \right)}\abs{\text{sin}\left(\frac{2\pi f}{f_L}\right)}\;,
\end{equation}
where $G=6.67\times10^{-11}\text{ m}^3\text{kg}^{-1}\text{s}^{-2}$ is the gravitational constant, $f_h=v_0/h$, $f_L=2v_0/L_{12}$ and $K_0$ is the zeroth order modified Bessel function of the second kind (see Appendix~\ref{apen:sigderiv} for a full derivation). This is purely range acceleration. For parameters relevant to the current GRACE-FO mission ($h\approx500$ km and $L_{12}\approx200$ km) this signal is approximately linear in $L_{12}$ in the low frequency limit.

\subsection{Measurement noise}
The measurement noise in this simple model comes in three forms, with two possible sources of accelerometer noise. Fig.~\ref{fig:scheme} shows a detailed schematic of the current GRACE-FO mission with the main noise sources highlighted in different colours. Non gravitational forces acting on the satellite contribute a non-gravitational phase shift to the laser light (highlighted in blue in Fig.~\ref{fig:scheme} a)). Thus, the total measured phase shift is $\phi(t)=\phi^{\text{g}}(t)+\phi^{\text{ng}}(t)$, where superscript (n)g denotes the phase shift due to (non-)gravitational forces. Before the range acceleration inferred from the measured phase can be compared to the expected acceleration based on the current best known gravitational field, the non-gravitational accelerations are removed using accelerometer data. This adds accelerometer instrument noise to the measurement with root power spectral density of the form (green box in Fig.~\ref{fig:scheme} b))
\begin{equation}
\label{an_aspec}
\sqrt{S_{\text{AN}}(f)}=\sqrt{2}a_0\sqrt{1+\left(\frac{f_k}{f}\right)^2}\;,
\end{equation}
where the maximum sensitivity of the accelerometer is defined by the acceleration white noise $a_0$ and the low frequency noise of the accelerometer is defined by $f_k=5\text{ mHz}$~\cite{touboul1999electrostatic}. The current GRACE-FO mission has \mbox{$a_0\approx100\text{ pm s}^{-2}\sqrt{\text{Hz}}^{-1}$} and it is anticipated that the next generation of GRACE will have \mbox{$a_0\approx1\text{ pm s}^{-2}\sqrt{\text{Hz}}^{-1}$}~\cite{spero2021point}. Note however, that this is only an approximate model for the accelerometer instrument noise on the GRACE-FO mission.

The LRI measures the phase between the two satellites with sub-micrometer precision, however there is some laser phase noise remaining in this measurement. The current instrument makes two measurements, one on each satellite, which are combined into a single useful measurement. After removing the non-gravitational element the remaining signal is
\begin{equation}
\label{eq:maintxtlpn}
2\hat{\phi}^{\text{g}}_{12}(t)=2\phi^{\text{g}}_{12}(t)+C_1(t-2\tau_{12})-C_1(t)+N_{12}(t)\;,
\end{equation}
%
%\begin{equation}
%\hat{\phi}^{\text{g}}_{21}(t)=\phi^{\text{g}}_{21}(t)+C_1(t-\tau_{12})-C_2(t)+N_{21}(t)\;,
%\end{equation}
where $\hat{\phi}^{\text{g}}_{ij}$ denotes the estimate of the gravitational phase shift measured at satellite $i$ using the light arriving from satellite $j$, $\tau_{ij}$ is the single-trip time of flight for light along that arm, $C_i(t)$ denotes the phase noise of the laser at satellite $i$ at time $t$ and $N_{ij}(t)$ denotes other noise sources in the measurement of light arriving at satellite $i$ from satellite $j$ (i.e. accelerometer instrument noise and quantum noise). For small $\tau_{12}$ ($L_{12}/c\ll1$),  this implies that the laser phase noise is proportional to satellite separation, as discussed in Appendix~\ref{lpnScaleLin}. For increased laser stability, one of the lasers is locked to an optical cavity and the second laser is then locked to the first. The current GRACE-FO mission requirement on the laser phase noise (red box in Fig.~\ref{fig:scheme} b)) has the following form
\begin{equation}
\begin{split}
\label{lpn_aspec}
\sqrt{S_{\text{LPN}}(f)}<&x_c\sqrt{1+\left(\frac{3\text{ mHz}}{f}\right)^2}\\
&\times\sqrt{1+\left(\frac{10\text{ mHz}}{f}\right)^2}\left(\frac{L_{12}}{220\text{ km}}\right)(2\pi f)^2\;,
\end{split}
\end{equation}
where $x_c\approx 80\text{ nm}\sqrt{\text{Hz}}^{-1}$~\cite{abich2019orbit} is a constant which we call laser white noise. However, the actual mission performance of the optical cavity exceeded this requirement. The actual laser phase noise performance is at the level of the cavity thermal noise
\begin{equation}
\label{lpn_aspec2}
\sqrt{S_{\text{LPN}}(f)}=\frac{(2\pi f)^2x_TL_{12}}{\sqrt{f}}\;,
\end{equation}
where $x_T\approx 1\times 10^{-15}$ is a constant which we call laser thermal noise~\cite{spero2021point}.

Both satellites in the GRACE-FO mission have a photoreceiver to measure the incoming light and fluctuations in the received photon number manifest as quantum noise. The quantum noise spectrum has the following form (orange box in Fig.~\ref{fig:scheme} b))
\begin{equation}
\label{sn_aspec}
\sqrt{S_{\text{QN}}(f)}=\sqrt{2}(2\pi f)^2\delta_{QN}\;,
\end{equation}
where $\delta_{\text{QN}}$ is a factor dependent on the amount of received power (discussed in more detail in Appendix~\ref{sec:loss}) and the factor of $\sqrt{2}$ comes from the fact that two measurements are made. A received power of $1$ nW corresponds to a quantum noise level of \mbox{$\delta_{\text{QN}}\approx1$ pm$\sqrt{\text{Hz}}^{-1}$} (note that this assumes homodyne detection and near-unity detection efficiency). Quantum noise is not presently a limiting factor, but it may be once other sources of noise are addressed and the interferometer becomes quantum-limited. These signal and noise spectra allow a complete characterisation of this model and are summarised in Fig.~\ref{fig:noise_spec} a). 

%\section{Time delay interferometry applied to a GRACE-like mission.}

\begin{figure*}[!ht]
\includegraphics[width=\textwidth]{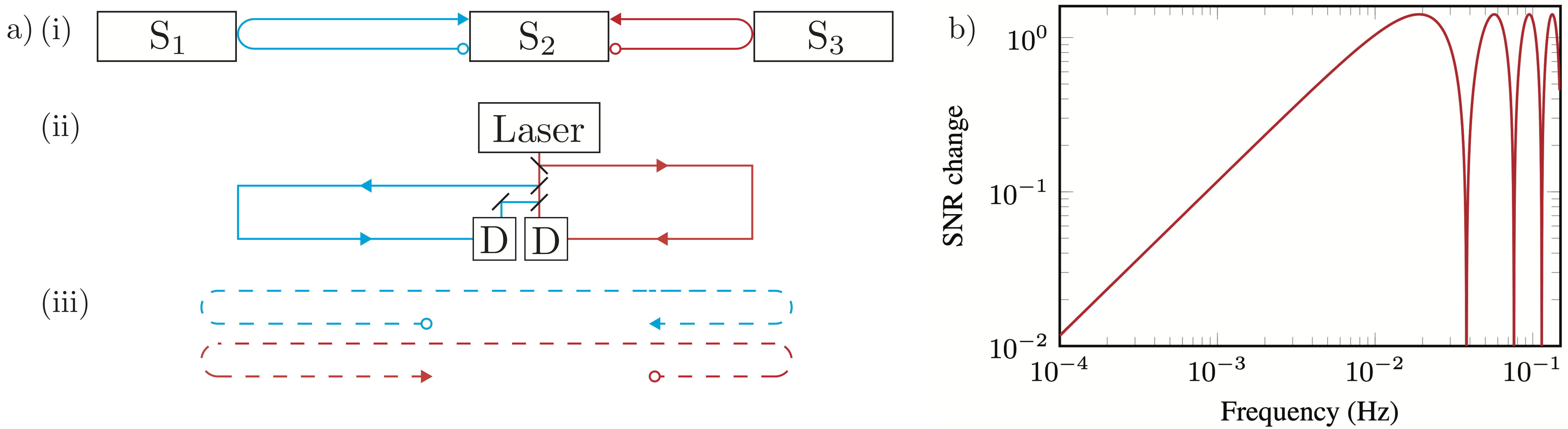}
\caption{\textbf{TDI applied to GRACE-like mission}. a) Schematic showing how the different length optical paths are converted to optical paths of the same effective length through TDI. (i) shows the true optical paths between the three satellites. (ii) shows how these optical paths are measured with the same laser. Finally (iii) shows how the effective optical path lengths for the two beams are equal after TDI. D represents the detection process, either homodyne or heterodyne detection. The different colours for the optical paths are for illustrative purposes only. b) Ratio of the SNR with TDI to the SNR without TDI for both quantum noise and accelerometer noise assuming a satellite velocity of $v_0=7600$ m/s and satellite separation of $L_{12}=200$ km. There are certain frequencies where the SNR is enhanced.}
\label{fig:tdisuper}
\end{figure*}

\subsection{Time delay interferometry for geodesy}
We now show how TDI can be used to significantly reduce laser phase noise (Eqs.~\eqref{lpn_aspec}, \eqref{lpn_aspec2}). TDI is a post-processing technique that uses multiple measurements, recombined with different time offsets, to cancel out common-mode noise~\cite{tinto1999cancellation}. To this end we consider multiple satellite formations with several measurements being made. Formation $\alpha^T_3$, from Fig.~\ref{fig:multscheme} b), with a single laser on the middle satellite is examined in detail, however many other combinations are possible, including combinations with multiple lasers, discussed in Appendix~\ref{apen:altlpnfree}. For formation $\alpha^T_3$ the middle satellite acts as the master satellite for the fleet. Light is split into four paths using beamsplitters, with two light beams being sent to the two outer satellites, where they are reflected back to the middle satellite (in practice this would be implemented using phase locked loops and second lasers, as shown in Fig.~\ref{fig:scheme} b), rather than mirrors). At the middle satellite two independent measurements are made, using the two light beams which remained on the middle satellite as local oscillators, shown in Fig.~\ref{fig:tdisuper} a). A similar TDI combination has been considered before for detecting gravitational waves~\cite{estabrook2003syzygy}. After removing the non-gravitational phase shift from the measurement using accelerometer data at time $t$ ($a(t)=\ddot{\phi}(t)\lambda/2\pi$) the measured signal is
\begin{equation}
2\hat{\phi}^{\text{g}}_{21}(t)=2\phi^{\text{g}}_{21}(t)+C_2(t-2\tau_{21})-C_2(t)+N_{21}(t)\;,
\end{equation}
\begin{equation}
2\hat{\phi}^{\text{g}}_{23}(t)=2\phi^{\text{g}}_{23}(t)+C_2(t-2\tau_{23})-C_2(t)+N_{23}(t)\;,
\end{equation}
using the same notation as before. %The extra factor of 2 comes from the fact that the light now travels along the path between the two satellites twice.
In order to cancel out the laser phase noise the effective optical path length needs to be the same for both beams, as is illustrated in Fig.~\ref{fig:tdisuper} a). The following combination of the blue (LHS) and red (RHS) optical paths achieves this:
\begin{equation}
\label{TDIcomb}
\begin{split}
&\text{ }2([\hat{\phi}^{\text{g}}_{21}(t)-\hat{\phi}^{\text{g}}_{23}(t)]-[\hat{\phi}^{\text{g}}_{21}(t-2\tau_{23})-\hat{\phi}^{\text{g}}_{23}(t-2\tau_{21})]) \\ 
=&\text{ }2[\phi^{\text{g}}_{21}(t)-\phi^{\text{g}}_{21}(t-2\tau_{23})+\phi^{\text{g}}_{23}(t-2\tau_{21})-\phi^{\text{g}}_{23}(t)]+\\
&\text{ }(N_{21}(t)-N_{21}(t-2\tau_{23}))-(N_{23}(t)-N_{23}(t-2\tau_{21}))\;.
\end{split}
\end{equation}

Converting to the frequency domain gives a signal which can be compared to the original scheme where TDI was not employed, $\alpha^O_2$. In order to do so we make the simplification that both satellite separations are initially equal, $L_{12}=L_{23}=L$. Each $\phi^{\text{g}}_{ij}$ term corresponds to the differential acceleration of one pair of satellites (Eq.~\eqref{eq:acc_twosat}). $\phi^{\text{g}}_{23}(t)$ is then equal to $\phi^{\text{g}}_{21}(t)$, delayed by the time-period ($v_0/L$). The TDI signal includes an additional delay on each $\phi^{\text{g}}_{ij}$ term, but by a time-period corresponding to the time of flight of the light (the previous delay corresponded to the time of flight of the satellite). This gives the signal in the frequency domain after TDI, as:
 %This is then delayed again by the same time-period ($v_0/L$) to convert $\phi^{\text{g}}_{21}(t)$ to $\phi^{\text{g}}_{23}(t)$.
 %Finally this is delayed a third time, but by a time-period corresponding to the time of flight of the light (previous delays corresponded to the time of flight of the satellite). This gives the signal in the frequency domain after TDI, as:
\begin{equation}
\begin{split}
 \label{eq:acc_TDIsat}
 \hspace{-1cm}
\abs{a_{\text{R,TDI}}(f)}=&\text{ }\frac{64\pi f GM}{v_0^2}\abs{K_0\left( \frac{2\pi f}{f_h} \right)}\\
&\times\abs{\text{sin}\left(\frac{2\pi f}{f_L}\right)^2}\abs{\text{sin}\left(\frac{\pi f}{f_c}\right)}\;,
\end{split}
\end{equation}
where $f_c=c/2L$, see Appendix~\ref{apen:gravsigtdi} for more detail. Clearly if the distance along the two arms is the same, TDI is not necessary as the measured signals can simply be subtracted with no time delay to remove the laser phase noise. However, in practice all three satellites will fly along slightly different trajectories and experience different non-gravitational accelerations. Therefore, even if the satellites are approximately evenly spaced, TDI will still be necessary to cancel the laser phase noise. In Appendix~\ref{apen:unequalsep} we consider the signal after using TDI when the two arm lengths ($L_{12}$ and $L_{23}$) are different, however, this does not significantly affect our results.

\textit{Signal-to-noise ratio after TDI.} Although laser phase noise can in principle be completely cancelled, imperfections in our knowledge of the satellite positions will hinder how well the laser phase noise is suppressed. In Appendix~\ref{LPNexp} we show that the noise spectrum of the laser phase noise after TDI, $\sqrt{S_{\text{LPN}_{\text{left-over}}}(f)}$, is approximately given by
\begin{equation}
\sqrt{S_{\text{LPN}_{\text{left-over}}}(f)} \approx8\pi\delta f\sqrt{S_{\text{LPN}}(f)} \;,
\end{equation}
where $\delta$ is the error in how well the time of flight for light between the two satellites is known. With GPS satellite positioning on the order of 5 mm~\cite{kroes2005precise,wu2006real}, $\delta=5\text{ mm}/c\approx10^{-10}\text{ s}$. Hence, at frequencies close to $10^{-2}$ Hz, the laser phase noise can be cancelled by approximately 10 orders of magnitude. This is more than sufficient to ensure the laser phase noise is no longer a dominant noise source. However, the signal-to-noise ratio (SNR) of the remaining noise sources is influenced by TDI. As a result of applying TDI the signal is affected such that
\begin{equation}
\frac{\abs{a_{\text{R,TDI}}(f)}}{\abs{a_{\text{R}}(f)}}=4\abs{\text{sin}\left(\frac{2\pi f}{f_L}\right)}\abs{\text{sin}\left(\frac{\pi f}{f_c}\right)}\,.
\end{equation}
Similarly, the remaining noise sources are affected in the following manner
\begin{equation}
\frac{\sqrt{S_{\text{AN,TDI}}(f)}}{\sqrt{S_{\text{AN}}(f)}}=2\sqrt{2}\abs{\text{sin}\left(\frac{\pi f}{f_c}\right)}\;,
\end{equation}
and
\begin{equation}
\frac{\sqrt{S_{\text{QN,TDI}}(f)}}{\sqrt{S_{\text{QN}}(f)}}=2\sqrt{2}\abs{\text{sin}\left(\frac{\pi f}{f_c}\right)}\;,
\end{equation}
as discussed in Appendix~\ref{tdi_noise_sources_apen}. Thus, the ratio of the SNR with TDI to the SNR without TDI for both the quantum noise and the accelerometer noise is $\sqrt{2}\abs{\text{sin}\left(2\pi f/f_L\right)}$. The change in SNR as a function of frequency for both quantum noise and accelerometer noise is shown in Fig.~\ref{fig:tdisuper} b). Importantly, the SNR for the remaining noise sources is enhanced by $\sqrt{2}$ when $f=v_0/2 L\approx1\times10^{-2}$ Hz, which is very close to the frequency of interest where the point mass gravitational signal is maximal. The reason for the SNR enhancement is that we are now measuring the phase shift between two pairs of satellites, instead of one pair as in the original mission. However, the SNR is degraded at certain frequencies, near the nodes in Fig.~\ref{fig:tdisuper} b), and one important implication of this is that TDI is most beneficial when the laser phase noise is the dominant noise source. A comparison of the signal and total noise spectra, both with and without TDI, is shown in Fig.~\ref{fig:noise_spec}. This shows the SNR enhancement that TDI can offer over a current GRACE style mission. Note that for real satellite geodesy missions, the frequencies of interest cover a broad range, at some of which, TDI will degrade the SNR with respect to the remaining noises. A more complete analysis will be required to determine the utility of TDI for real world satellite geodesy.

\begin{figure}[t]
\includegraphics[width=0.4\textwidth]{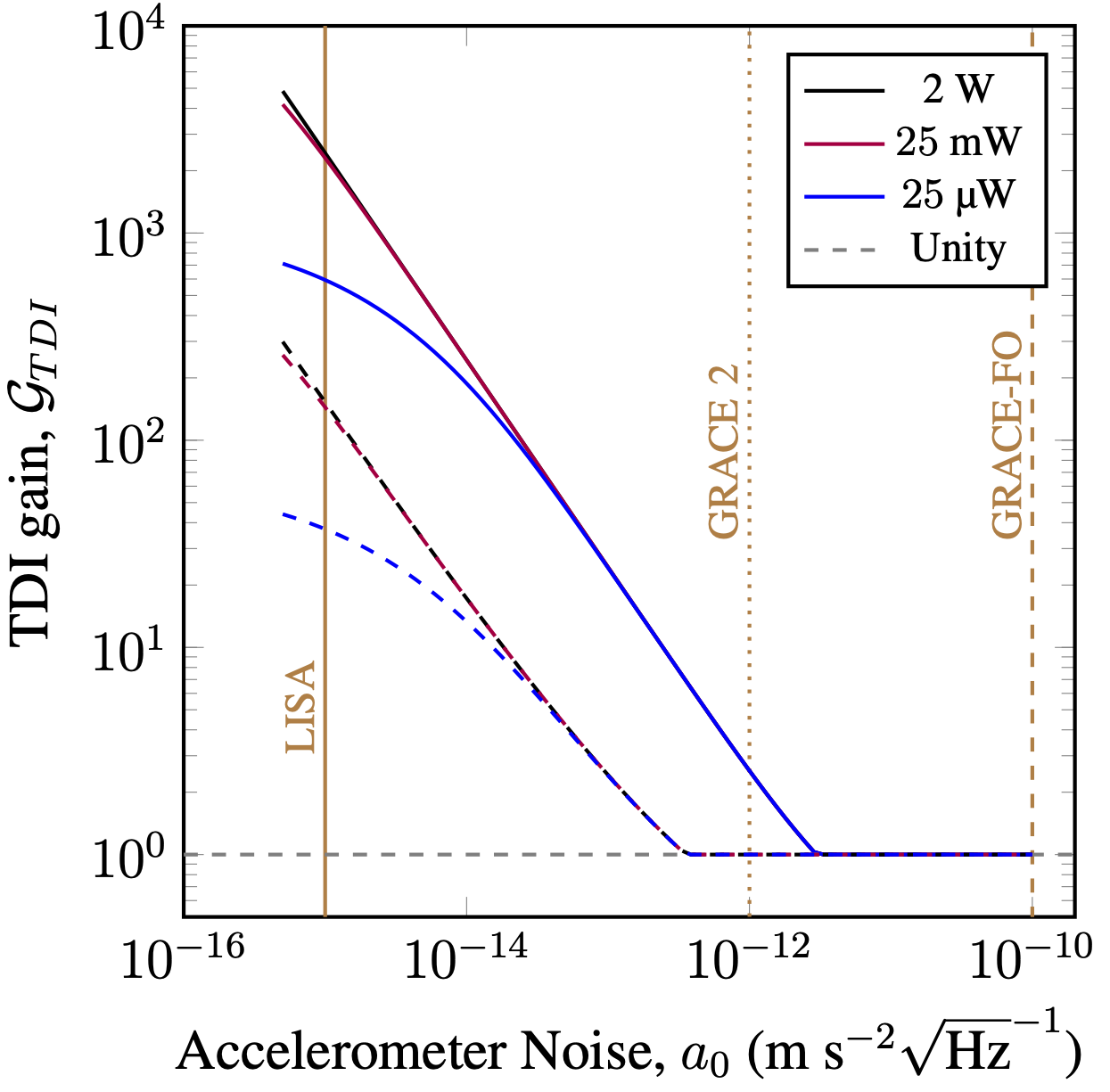}
\caption{\textbf{Point mass sensitivity enhancement due to TDI.} Plotted is the TDI gain as a function of accelerometer instrument noise for the two different laser phase noise performances, the GRACE-FO laser phase noise requirement, Eq.~\eqref{lpn_aspec} (solid lines, $x_c=8\text{ nm}\sqrt{\text{Hz}}^{-1}$) and the actual laser phase noise performance, Eq.~\eqref{lpn_aspec2} (dashed lines, $x_T= 1\times 10^{-15}$). The TDI gain increases with decreasing accelerometer instrument noise as at lower accelerometer instrument noise levels the cancellation of the laser phase noise is more impactful. However, this does not increase indefinitely as eventually the quantum noise limit is reached. Three different quantum noise levels are considered, corresponding to transmitted powers of 2 W, 25 mW and 25 $\mu$W, at a satellite separation of $L_{12}=L_{23}=200$ km with a receiving aperture radius of 5 cm. The satellite orbital height is $h=500$ km. The vertical dashed, dotted and solid brown lines show the projected accelerometer instrument noise for the GRACE-FO mission, the next GRACE mission (GRACE 2) and the LISA mission respectively. At large accelerometer noises the TDI gain is never less than 1, because TDI is a non-destructive measurement.}
\label{fig4new}
\end{figure}

\begin{figure}[t]
\includegraphics[width=0.45\textwidth]{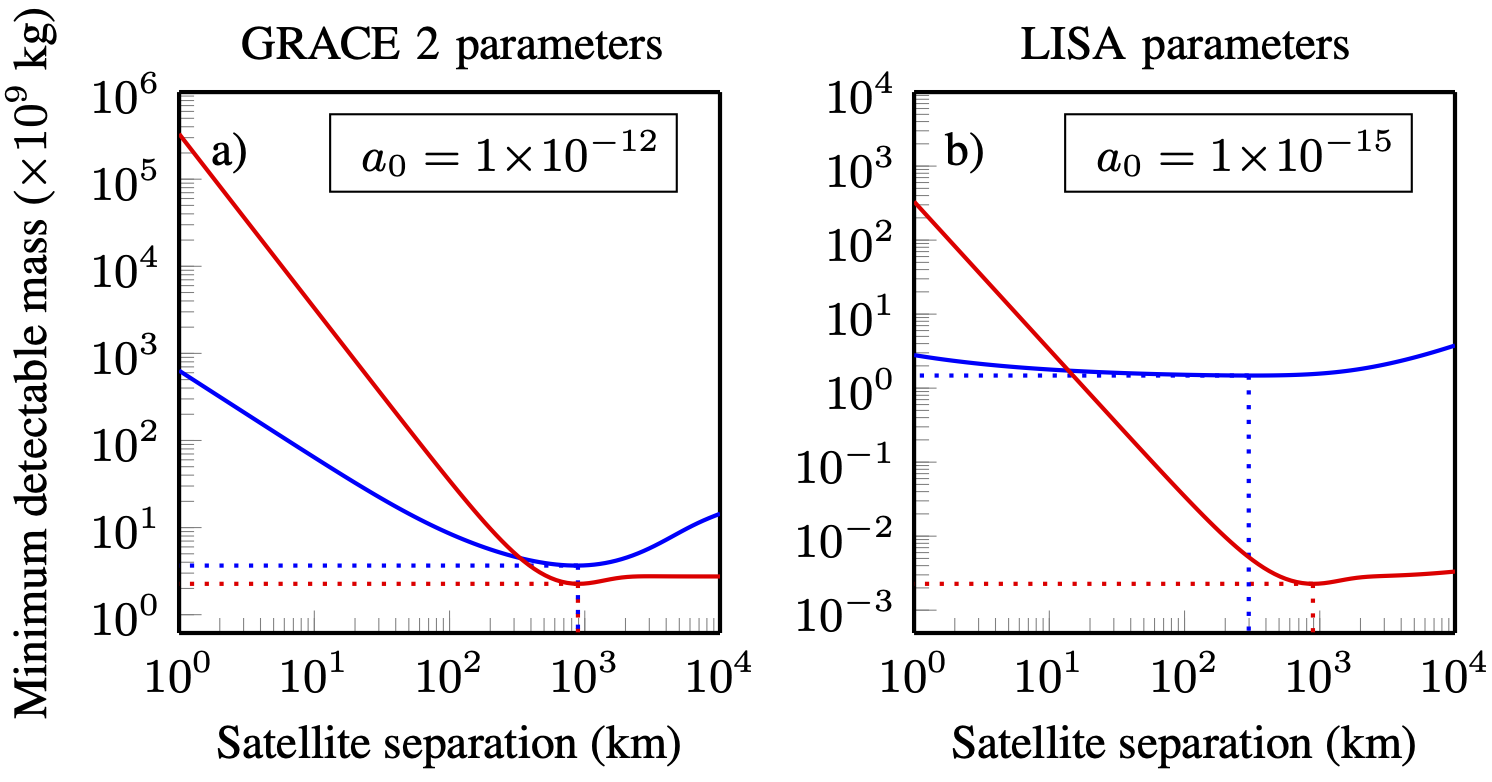}
\caption{\textbf{Optimal satellite separation for point mass sensitivity}. Shown is the point mass sensitivity both with (red lines) and without (blue lines) TDI for accelerometer noises of $a_0=1\times10^{-12}\text{m }\text{s}^{-2}\sqrt{\text{Hz}}^{-1}$ (a) and $a_0=1\times10^{-15}\text{m }\text{s}^{-2}\sqrt{\text{Hz}}^{-1}$ (b) as a function of the satellite separation. Laser thermal noise is $x_T= 1\times 10^{-15}$ and satellite orbital height is $h=500$ km. The optimal satellite separation is that which minimises the minimum detectable mass, and is different for different mission configurations.}
\label{fig5new}
\end{figure}

\textit{Minimum Detectable Mass.} It is to be expected that TDI can aid satellite geodesy as one of the major noise sources is removed without the signal being totally compressed. This can be made rigorous by considering the minimum detectable mass defined as~\cite{spero2021point}
\begin{equation}
M_{\text{min}}=\frac{3}{\sqrt{4\int_{0}^{\infty}\frac{\abs{a_{\text{R}}(f)/M}^2}{S_{\text{T}}(f)}df}}\;,
\label{minmass}
\end{equation}
where $S_{\text{T}}$ is the total noise spectrum given by
\begin{equation}
S_{\text{T}}=S_{\text{AN}}+S_{\text{LPN}}+S_{\text{QN}}\;.
\end{equation}
This is the minimum mass which corresponds to a SNR of at least 3, which intuitively represents the smallest possible mass our system can detect. We now define the following quantity as the TDI gain
\begin{equation}
\mathcal{G}_{TDI}=\frac{M_{\text{original}}}{M_{\text{TDI}}}\;,
\end{equation}
where $M_{\text{original}}$ is the minimum detectable mass in the original scheme without TDI, formation $\alpha^O_2$, and $M_{\text{TDI}}$ is the minimum detectable mass with TDI, formation $\alpha^T_3$. Intuitively the TDI gain tells us how many times smaller a mass can be detected with TDI than without. 

With realistic future accelerometer instrument noise levels, TDI has the potential to significantly reduce the minimum detectable mass, as shown in Fig.~\ref{fig4new}. At high accelerometer instrument noises the laser phase noise is not important and so TDI does not offer any improvement. However, as TDI is a non-destructive measurement we can simply choose not to use TDI in postprocessing. With reducing accelerometer instrument noise the TDI gain increases, until when the accelerometer instrument noise is sufficiently low, quantum noise becomes the major noise source and so the advantage flattens off. At very low values of $a_0$, when quantum noise starts to dominate there is an advantage to increasing the laser power. Equivalently, this advantage can be obtained from increasing the receiving aperture size, or any technique to reduce quantum noise, such as optical squeezing~\cite{schnabel2010quantum}. For sufficiently small accelerometer and quantum noise levels, the left-over laser phase noise after TDI may become the limiting factor again. The sensitivity gain offered by TDI is very close to being achievable with today's technology, with the accelerometers of past and planned missions having $a_0$ values in the range $a_0\approx 1\times10^{-12}\text{ m}\text{ s}^{-2}\sqrt{\text{Hz}}^{-1}\rightarrow a_0\approx 1\times10^{-15}\text{ m}\text{ s}^{-2}\sqrt{\text{Hz}}^{-1}$~\cite{armano2018beyond,christophe2010orbit,alvarez2021simplified}.

%for the LISA mission projected to have $a_0\approx 1\times10^{-15}\text{ m}/\text{s}^2\sqrt{\text{Hz}}$~\cite{armano2018beyond}. fac 500 improvement~\cite{alvarez2021simplified} goce~\cite{}

In Fig.~\ref{fig5new} the minimum detectable mass as a function of satellite separation is shown for geodesy both with and without TDI. The optimal satellite separation (that which minimises the minimum detectable mass) differs depending on the strategy employed. Without TDI, improvements in the accelerometer instrument noise produce only marginal improvements in sensitivity. However, the same improvement combined with TDI can vastly improve sensitivity. Without TDI, the upgrade of the GRACE-FO accelerometer from $a_0 = 1\times10^{-12}\text{ m }\text{s}^{-2}\sqrt{\text{Hz}}^{-1}$ to $a_0 = 1\times10^{-15}\text{ m }\text{s}^{-2}\sqrt{\text{Hz}}^{-1}$ induces a very minor improvement. In contrast, the use of TDI in the same conditions leads to an improvement of nearly three orders of magnitude, with the minimum detectable mass being almost $1\times10^{6}$ kg. For perspective, this mass is equivalent to a change in water or ice levels almost as small as 1 mm over a 1 km$^2$ area. However, we note again that these calculations are based on the 1D point mass model and so are not directly related to actual satellite geodesy missions. Additionally, if the laser phase noise and accelerometer noise are sufficiently reduced, other noise sources may start to dominate~\cite{dobslaw2016modeling,wegener2020tilt}. In Appendix~\ref{MDM_apen} similar calculations are presented for a range of satellite orbital heights.

The current GRACE-FO mission uses an optical cavity to achieve an improved frequency stability. We now compare the point mass sensitivity, both with and without TDI, in terms of requisite laser stability. The leftover laser phase noise is calculated assuming the satellite positions are known to within 5 mm. For $a_0=2\times10^{-13}\text{ m }\text{s}^{-2}\sqrt{\text{Hz}}^{-1}$, to achieve the same point mass sensitivity as is provided by using TDI and a laser with $x_T\approx1\times10^{-12}$, without using TDI requires a laser with three orders of magnitude more stability, $x_T=1\times10^{-15}$. However, as above, for real-world geodesy this relaxation in laser stability may not be true owing to the more complex frequency dependence of the gravitational signal. Specifically, for recovering signals at low frequencies where TDI degrades the SNR, this relaxation in laser stability would not be possible.
%Thus TDI can allow a considerable saving of time, money and weight with regards the optical cavity.

\subsection{Accelerometer Noise}
The purpose of the accelerometer is to measure the non-gravitational acceleration as accurately as possible so that it can be removed from the measurement while adding the minimum amount of noise. Ultimately however, the accelerometer will always add some noise. We now show that through precise satellite engineering and formation flying, the line-of-sight non-gravitational acceleration can be removed from the measurement without using an accelerometer. This eliminates a major noise source, accelerometer instrument noise. The principle behind this is that the non-gravitational forces acting on the satellites consist of a stationary and a non-stationary component. These non-gravitational forces then give rise to non-gravitational accelerations, which have a stationary, $a^{\text{ng}}_{\text{s}}$ and a non-stationary, $a^{\text{ng}}_{\text{ns}}$, component. Stationarity here refers to temporal stationarity. If the leading satellite is at position $x$ at time $t$ and the trailing satellite reaches $x$ at a time $t+\Delta t$, then the stationary non-gravitational accelerations will be common to both satellites, $a^{\text{ng}}_{\text{s}}(x,t)=a^{\text{ng}}_{\text{s}}(x,t+\Delta t)$ and the non-stationary non-gravitational forces will differ $a^{\text{ng}}_{\text{ns}}(x,t)\neq a^{\text{ng}}_{\text{ns}}(x,t+\Delta t)$. 

The stationary non-gravitational accelerations experienced by all satellites will be the same provided they have the same mass and identical aerodynamicity. The similarity of the non-stationary non-gravitational accelerations experienced by each satellite, i.e. how much $a^{\text{ng}}_{\text{ns}}(x,t)$ and $a^{\text{ng}}_{\text{ns}}(x,t+\Delta t)$ differ, is correlated to the satellite separation. The further the satellites are apart the more the non-stationary component will have changed by the time it takes the trailing satellite to reach the position of the leading satellite.

\textit{Six satellite differential mass formation flying with TDI.} We now turn to the satellite combinations presented in Fig.~\ref{fig:multscheme} c) and d), formations $\alpha^T_{3,DM}$ and $\alpha^T_{6,DM}$ respectively. Neither of these combinations require an accelerometer and so do not introduce any accelerometer instrument noise. Instead these formations rely on satellites of precisely known, but different masses which will experience different non-gravitational accelerations. Assuming identical aerodynamicity, the same non-gravitational force acting on two satellites, one with twice the mass of the other, will result in twice the non-gravitational acceleration for the lighter satellite. This principle allows common mode non-gravitational accelerations to be removed from the measurement, and is the reason an accelerometer is no longer required. Formation $\alpha^T_{3,DM}$ relies on only three satellites separated by distances on the order of hundreds of kilometers. As the satellites are so distant from one another, only the stationary component of the non-gravitational accelerations will be common to all three satellites, allowing this to be removed from the measurement. This formation does not allow the non-stationary component of the non-gravitational acceleration to be removed. As such, formation $\alpha^T_{3,DM}$ performs worse than the current GRACE-FO mission with realistic parameters and so we defer further discussion of this to Appendix~\ref{AltAccComb}.
\begin{figure*}[t]
\centering
\includegraphics[width=\textwidth]{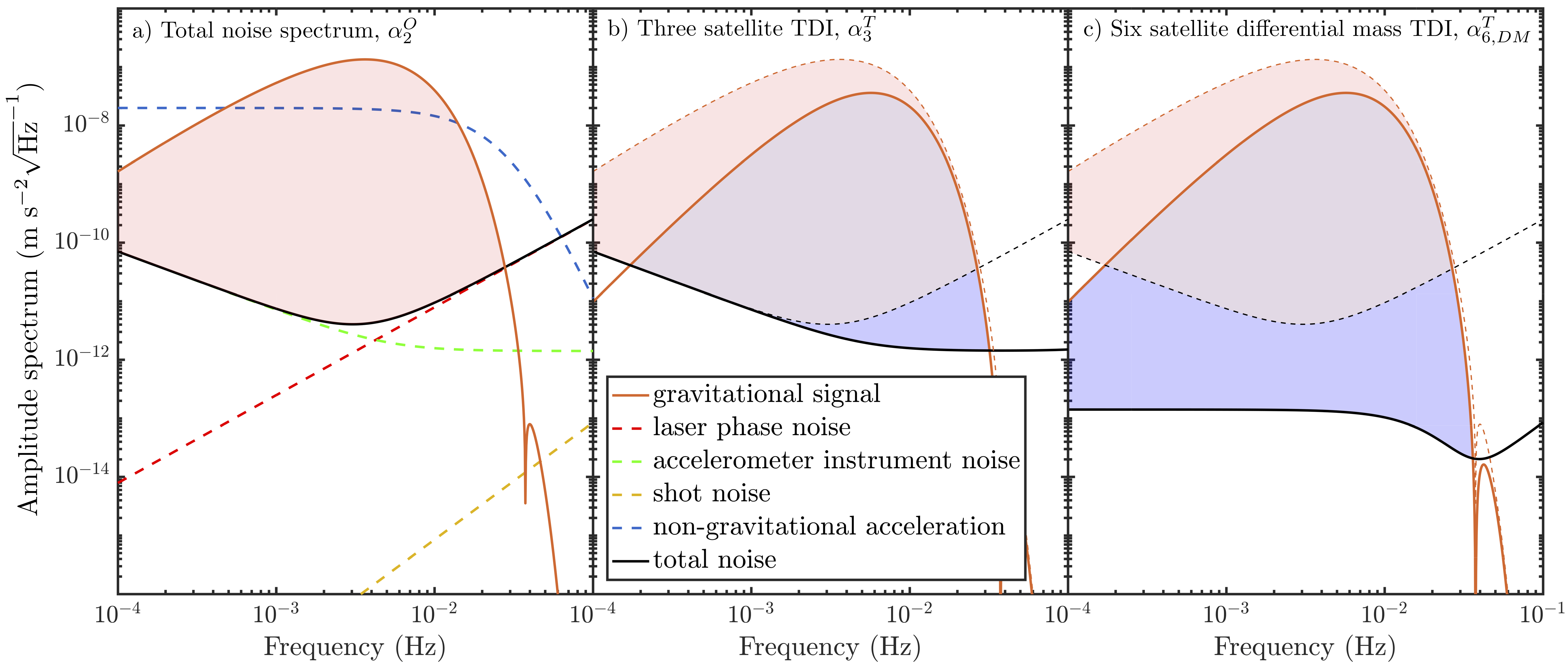}
\caption{\textbf{Noise spectrum analysis of proposed satellite geodesy missions.} a) Signal and noise spectra for a GRACE-FO-like, two-satellite mission, formation $\alpha^O_2$. The gravitational signal corresponds to a $1\times10^{13}$ kg point mass, satellite separation of $L=$ 200 km and satellite orbital height of $h=$ 500 km (Eq.~\eqref{eq:acc_twosat}), accelerometer instrument noise corresponds to $a_0=1\times10^{-12}\text{ m }\text{s}^{-2}\sqrt{\text{Hz}}^{-1}$ (Eq.~\eqref{an_aspec}), laser phase noise corresponds to $x_T=1\times10^{-15}$ (Eq.~\eqref{lpn_aspec2}), the quantum noise level is calculated by considering the diffraction limits set by a 25 cm receiving aperture radius and 25 mW of initial optical power (Eq.~\eqref{sn_aspec}) and the non-gravitational accelerations are those from Eq.~\eqref{NGfreqspec}. The region highlighted in red corresponds to frequencies where the signal is above the noise floor, i.e. the region which contributes most to enhancing the signal-to-noise ratio. b) Signal and total noise spectrum for a formation $\alpha^T_3$ mission, with three satellites and TDI being employed. The gravitational signal after TDI is given by Eq.~\eqref{eq:acc_TDIsat}. The dashed orange and black lines correspond to the gravitational signal and total noise from the $\alpha^O_2$ mission. The region highlighted in red corresponds to regions which can only be accessed by the $\alpha^O_2$ mission, the region highlighted in blue shows the new region which can be accessed by the $\alpha^T_3$ mission and the mauve region in between is accessible for both missions. As TDI is a non-destructive measurement the region where the SNR > 1 for the $\alpha^O_2$ mission is still accessible. The increase in the size of the shaded region highlights the benefit of TDI in this instance. Satellite positions are assumed to be known to within 5 mm. The spectra corresponding to scheme $\alpha^T_3$ with TDI are rescaled by $1/(2\sqrt{2}\abs{\text{sin}(\pi f/f_c)})$ so that quantum noise and accelerometer noise are unaffected by TDI. c) Signal and total noise spectrum for a formation $\alpha^T_{6,DM}$ style mission, with six different mass satellites and TDI being employed. Again the blue region corresponds to the benefit of this scheme, the region which cannot be accessed by the $\alpha^O_2$ mission. Each A-B satellite pair is assumed to fly within 1 m of each other. The spectra corresponding to scheme $\alpha^T_{6,DM}$ with TDI are rescaled by $1/(2\sqrt{2}\abs{\text{sin}(\pi f/f_c)})$. For all of the plots the gravitational signals have units$\text{ m }\text{s}^{-2}\text{Hz}^{-1}$.}
\label{fig:noise_spec}
\end{figure*}

Formation $\alpha^T_{6,DM}$ is more promising as it is, in theory, able to \textit{completely remove laser phase noise and accelerometer noise.} This scheme works by making two independent sets of measurements with effectively the same laser. The scheme is broken into 3 pairs of different mass satellites, where, as before the different pairs will be separated by hundreds of kilometers. However, the satellites within each pair are required to stay as close as possible to each other. The two satellites in each pair, which are of mass M$_S$ and 2M$_S$, are called A and B satellites respectively. Owing to the different masses, the B satellites will experience half the non-gravitational accelerations the A satellites experience. Importantly, as each A-B pair is close to each other, they will experience almost the same stationary and non-stationary non-gravitational forces. This allows for the near-perfect removal of non-gravitational accelerations. Owing to the different non-gravitational accelerations experienced, thruster movements will be required to keep each pair close to each other. It is only by having two satellites with different masses close to each other that the non-stationary component of the non-gravitational accelerations can be removed. 

Satellites in this scheme are denoted $S_{i,j}$, with $i\in\{A, B\}$ denoting whether the satellite is the heavier (B) or lighter (A) of this particular pair and $j\in\{1,2,3\}$ denoting which pair of satellites we refer to (1 being the leading satellite and 3 the trailing satellite). The laser on satellite $S_{B2}$ is sent to satellite $S_{A2}$, and through a short delay fibre on satellite $S_{B2}$, it can be arranged that both satellites are using effectively the same laser. This light is sent to the outer satellites and reflected back to the middle satellites where two measurements are made by each of the middle A-B pair. Satellite $S_{A2}$ measures

\begin{equation}
%\centering
\begin{split}
2\hat{\phi}_{A21(3)}(t)=&\text{ }2\phi^{\text{g}}_{21(3)}(t)+C(t-2\tau_{21(3)})-C(t)\\ 
&+QN_{1(3)A}(t)-2\phi^{\text{ng}}_{1(2)}(t)+2\phi^{\text{ng}}_{2(3)}(t)\;,
\end{split}
\end{equation}
and 
$S_{2B}$ measures
\begin{equation}
\centering
\begin{split}
2\hat{\phi}_{B21(3)}(t)=&\text{ }2\phi^{\text{g}}_{21(3)}(t)+ C(t-2\tau_{21(3)})-C(t)\\
&+QN_{1(3)B}(t)-\phi^{\text{ng}}_{1(2)}(t)+\phi^{\text{ng}}_{2(3)}(t)\;,
\end{split}
\end{equation}
where $QN_{ij}(t)$ denotes the quantum noise at time $t$ on satellite $S_{j2}$ for light received from satellite $S_{ji}$. These measurements can be combined to give two total measurement terms with no accelerometer noise, $2\phi_{T21}(t)=4\hat{\phi}_{B21}(t)-2\hat{\phi}_{A21}(t)$ and $2\phi_{T23}(t)=4\hat{\phi}_{B23}(t)-2\hat{\phi}_{A23}(t)$. The laser phase noise can then be removed from these two measurements using the same TDI combination discussed earlier. We re-emphasise that in principle this combination requires \textit{no on-board accelerometer} and \textit{no optical cavity} provided the satellites can be flown with sufficient accuracy.

With an orbital height of $h=$ 500 km and satellite separation of $L=$ 200 km this scheme can achieve a minimum detectable mass of $9\times10^4$ kg, assuming perfect accelerometer noise cancellation, laser phase noise cancellation to within 5 mm and a transmitted laser power of 2 W. This is approximately 6 orders of magnitude better than the current GRACE mission ($3\times10^{11}$ kg, assuming $x_T=1\times10^{-15}$), more than 4 orders of magnitude better than the current mission with an improved accelerometer, $a_0=1\times10^{-12}\text{m }\text{s}^{-2}\sqrt{\text{Hz}}^{-1}$, ($5\times10^9$ kg) and approximately 3 orders of magnitude better than the TDI combination with an ambitious level of accelerometer instrument noise, $a_0=1\times10^{-14}\text{m }\text{s}^{-2}\sqrt{\text{Hz}}^{-1}$, (9$\times10^7$ kg). Note that the minimum detectable mass presented here for the current GRACE mission (\mbox{$3\times10^{11}$ kg}) is different to the value quoted by Spero~\cite{spero2021point}, as we use a slightly different definition of the signal strength and a different satellite separation. The potentially huge improvement in sensitivity makes the significant technological challenge of implementing this scheme one worth considering. 

In reality, the A-B satellites in each pair will not be in exactly the same position, and the separation of each pair will drift over time. The further apart the satellites in each A-B pair are, the larger the difference in the non-stationary component of the non-gravitational force experienced will be. The effect of this is that the non-gravitational acceleration cancellation will not be perfect. However, even if each pair of satellites cannot be made to fly exactly alongside one another, some cancellation can still be achieved. An approximate model for the difference in non-gravitational acceleration in the along-track direction experienced by a pair of satellites separated by 200 km, after data transplanting (a technique used to estimate the non-gravitational accelerations of one satellite using accelerometer data from the other satellite) is
\begin{equation}
a^{\text{ng}}(f, 200\text{ km})\approx\frac{a_n}{\left[1+\left(\frac{f}{f_n}\right)^2\right]^3}\;,
\label{NGfreqspec}
\end{equation}
where $a_n=2\times10^{-8}\text{ m s}^{-2}\sqrt{\text{Hz}}^{-1}$ and $f_n=3\times10^{-2}$ Hz~\cite{bandikova2019grace}. We make the assumption that the difference in non-gravitational accelerations will scale linearly with distance, such that $a^{\text{ng}}(f, x)=a^{\text{ng}}(f, 200\text{ km})\cdot\left(x/200\text{ km}\right)$. As the spectrum in Eq.~\eqref{NGfreqspec} is obtained when the non-gravitational data from one satellite has been transplanted, which is not the case with our scheme, the true differential non-gravitational accelerations will be larger than those predicted by this model. However, as the satellites become closer the difference between transplanting and not transplanting becomes smaller. In an ideal implementation of our scheme each satellite-pair will be separated by no more than a few meters, hence this difference would be small. After TDI the leftover non-gravitational accelerations are scaled by a factor of $2\sqrt{2}\abs{\text{sin}\left(\pi f/f_c\right)}$. This approximate model can be used to place bounds on the performance of this scheme. 

In order for this technique to outperform TDI alone this cancellation must be below the projected accelerometer noise. The non-gravitational acceleration can be further reduced if the overall drag of the satellites is reduced, as would be the case by transitioning to CubeSats, or changing orbital height. Imperfect satellite flying in this scheme means that the required TDI combination becomes slightly more complex, with the necessary combination shown in Appendix~\ref{SixSatApen}. If each A-B pair of satellites can be flown within 1 m of each other, while reducing the total drag of each satellite by a factor of ten compared to the current GRACE mission, then this technique for accelerometer noise cancellation is equivalent to having an accelerometer with $a_0\approx5\times10^{-15}\text{ m }\text{s}^{-2}\sqrt{\text{Hz}}^{-1}$ in terms of minimum detectable mass. Thus, although we are using a simplified model of the non-gravitational accelerations, it is possible that this scheme can yield significant improvements with future technologies. Fig.~\ref{fig:noise_spec} c) compares the signal and total noise spectra of a GRACE-FO-like mission in its present state with all noise sources to our proposed implementation of this six-satellite scheme, formation $\alpha^T_{6,DM}$, at an orbital height of 500 km, satellite separation of 200~km between trailing satellites and 1 m between each A-B pair of satellites. 

The enhancement discussed in this section relies on a number of simplifying assumptions which will not be true in practice, regarding the differential non-gravitational accelerations. For example, the satellite masses will change over the course of the mission, thruster firings will affect the satellite accelerations and the satellites won't have identical aerodynamicity. Our scheme is only capable of removing line-of-sight non-gravitational accelerations, hence for real world geodesy, accelerometers may still be necessary to remove 3D non-gravitational accelerations. Additionally, there are significant technological hurdles to overcome before a mission configuration as complex as this can be used in reality. Finally, there are many practical issues with flying each A-B satellite pair close to each other, such as the risk of collision and the fact that the satellites will drift apart and follow slightly different orbits. These imperfections will manifest as ranging errors. Nevertheless, with further development, the principle behind this configuration may one day be of great use to geodesy missions. For instance, it may be possible to avoid some of these difficulties by replacing each satellite pair with a single satellite containing two different mass test masses in freefall.

\subsection{Quantum Limited satellite geodesy}
\begin{figure}[t]
\includegraphics[width=0.45\textwidth]{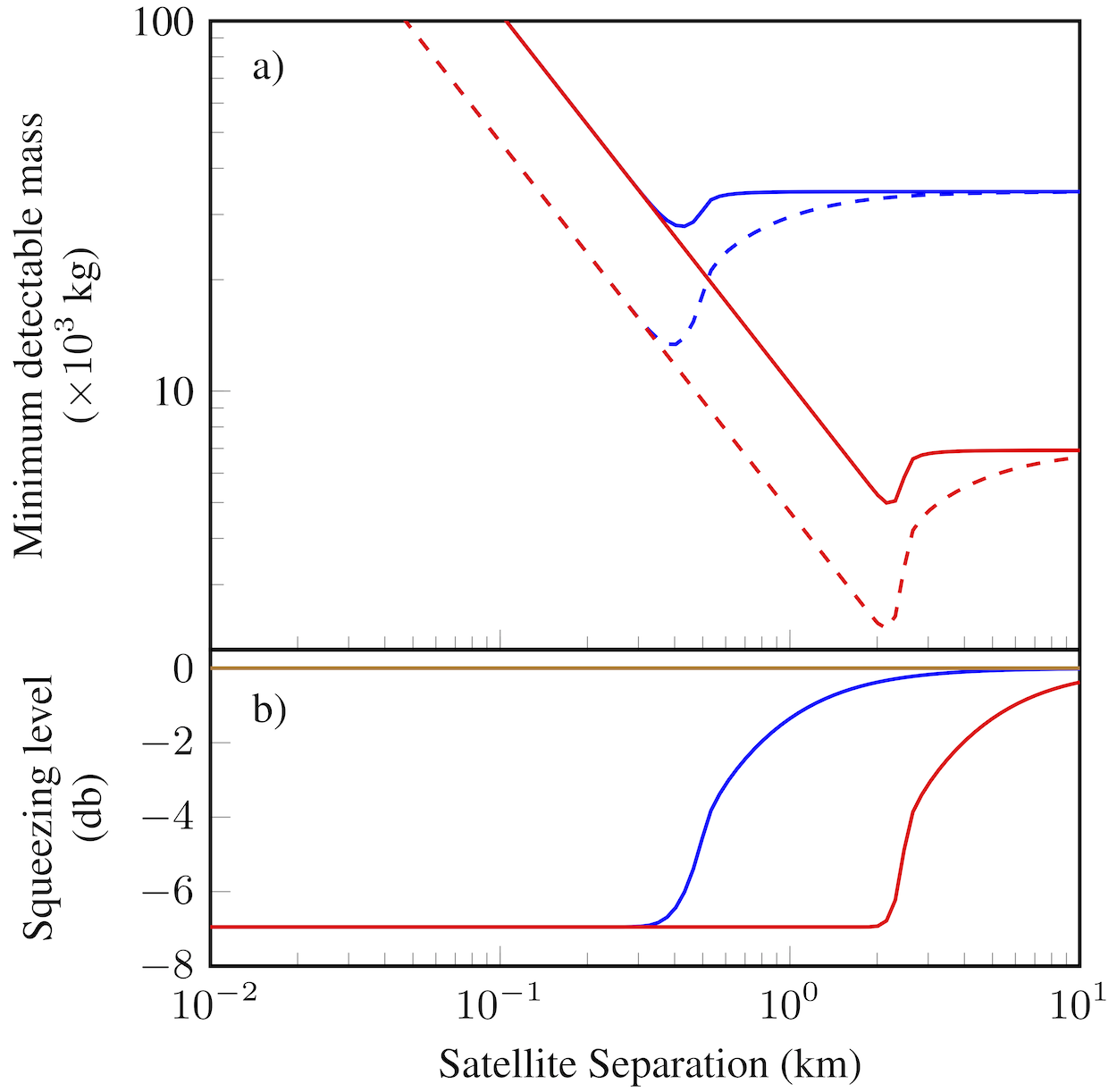}
\caption{\textbf{Minimum detectable mass for quantum noise limited geodesy.} a) The benefit from using squeezed light is obtained within the satellite separation where diffraction losses are not significant. This distance can be extended by increasing the aperture of the receiving optics. The blue and red lines correspond to receiving optics with aperture radius of 5 and 25 cm, respectively. Dashed lines indicate the point mass sensitivity using squeezed light. Parameters used are input power $P_0$=25 mW and initially 7 dB of pure squeezing. b) Effective squeezing level as a function of satellite separation. Brown line corresponds to the quantum noise level. }
\label{fignewreg}
\end{figure}
Having suggested schemes for reducing the laser phase noise and accelerometer noise, we now turn our attention to reducing the quantum noise limit of satellite-based geodesy. The current GRACE-FO mission is not quantum noise limited and so does not reach the fundamental quantum interferometry bound~\cite{demkowicz2013fundamental}. However, future satellite missions may one day approach the quantum limit. For example, the quantum noise limit can be reached either with instrument enhancement, i.e. improvements in optical cavity stability and accelerometer instrument noise, or through the multi-satellite formations presented above. One way to reduce quantum noise is using squeezed light~\cite{berni2015ab,goda2008quantum,xiao1987precision}, which would reduce the quantum noise term $\delta_{QN}$ in Eq.~\eqref{sn_aspec},  by a factor $e^{r}$, where $r$ is the squeezing level. Thus, from Eq.~\eqref{minmass} we can expect that squeezing can provide an enhancement of up to $e^{r}$ in terms of minimum detectable mass.

Interestingly, depending on how the quantum noise limited regime is reached, the optimal satellite separation is different. If the quantum noise limited regime is reached through enhancements in instrument noise there is a new regime which is optimal for satellite geodesy. When quantum noise limited, a larger satellite separation increases the signal strength but also increases the noise floor as the received optical power and squeezing level are reduced. The optimal satellite separation is that which minimises this trade-off, as shown in Fig.~\ref{fignewreg}. The smallest minimum detectable mass now occurs at the point where diffraction loss first becomes noticeable, which for 25 cm receiving apertures is at approximately 2 km. This is not at all obvious as at greater satellite separations the signal strength is much larger. However, by transitioning to a mission with reduced satellite separation, the benefits of squeezing and a greater received optical power compensate for the reduced signal strength. This was verified with a full 3D numerical simulation of satellites flying in the Earth's gravitational field when quantum noise limited, shown in Appendix~\ref{apen:phaseestSNL}. It should be noted that 25 cm radius receiving aperture optics would be considerably more expensive than what is presently used (for reference the LISA mission plans to use a 15 cm radius telescope~\cite{amaro2017laser}).

Alternatively, if the quantum noise limit is reached by multi-satellite formation flying combined with TDI, the optimal satellite separation can be much greater, extending far beyond the separation where squeezing stops being useful due to excessive propagation loss. This is because TDI will reduce the signal strength more as the two measurements become more correlated, which happens when the satellites are closer as the gravitational field experienced is more similar. This was also verified with our 3D model, discussed further in Appendix~\ref{apen:phaseestSNL}. In addition to the use of squeezing, several other quantum techniques, including optical delay lines and distributing multi-mode entangled states between satellites were investigated and found to have varying degrees of utility in the quantum noise limited regime. This will be the subject of future research.

For 200 km satellite separation, in order to be in the quantum noise limited region, significant technological progress is necessary, requiring a laser thermal noise of $x_T\approx 5\times10^{-20}$ and an acceleration white noise of $a_0\approx 2\times10^{-16}\text{ m s}^{-2}\sqrt{\text{Hz}}^{-1}$ as shown in Appendix~\ref{apen:SNLreq}. Furthermore, sub-hertz squeezing would be necessary if squeezed light is to be useful for geodesy. One might imagine that by transitioning to an alternate mission where the satellites are much closer (on the order of meters) we may enter a regime where quantum noise is the limiting factor as the laser phase noise will be greatly reduced. However, this is not the case as at smaller satellite separations the quantum noise is also greatly reduced due to the detection of more optical power. The high frequency roll-off in the gravitational signal makes the quantum noise limited regime difficult to reach for typical satellite parameters owing to the different frequency dependence of laser phase noise and quantum noise. The only way to get around the high frequency roll-off is by transitioning to lower orbital heights. In Appendix~\ref{apen:SNLreq} we propose a new type of mission which operates in this regime. Such a mission is impossible for mapping the Earth's gravitational field, but may find use for mapping the gravitational field of other astronomical bodies~\cite{gladstone2016atmosphere}. This type of mission appears to be in the quantum noise limited regime, allowing for squeezed light enhanced geodesy. However, quantum noise can also be further reduced by increasing the optical power. This pushes the need for squeezed light even further away. Increasing the optical power is currently a less technically challenging method of reducing the quantum noise than generating squeezed light in space. Nevertheless, the techniques presented in this section may someday be useful for satellite geodesy. We liken this to Carlton Caves' original proposal to use squeezed light in the search for gravitational waves~\cite{caves1981quantum}, which after decades of technological progress will reveal the quantum noise limit of an instrument, as in LIGO~\cite{aasi2013enhanced}.

\section{Discussion}
In this paper several techniques have been presented which can be applied to satellite geodesy to enhance the point mass sensitivity. The potential improvements we have proposed rely on some simplifying assumptions, namely circumscribing the analysis to point mass sensitivity and measurement noise. However, the proposed techniques show great promise which may motivate further studies with fewer assumptions. We have shown that time delay interferometry can offer significant benefits, in terms of the minimum detectable point mass. Time delay interferometry can be implemented with current technology and would be a useful technology demonstration for LISA. With a LISA-grade accelerometer, time delay interferometry can offer almost 3 orders of magnitude improvement in point mass sensitivity. Precisely controlled multi-satellite formations were presented which can remove accelerometer noise and laser phase noise simultaneously. Importantly, these formations do not require on-board accelerometers nor optical cavities. Finally, the possibility of reducing quantum noise through the injection of squeezing has been studied. Although squeezed light has the potential to improve satellite geodesy, significant technological enhancements are required before this becomes relevant. Nevertheless, we anticipate that the techniques presented here will have a crucial role to play in the enhancement of satellite geodesy in the future. 

\section*{Data availability}
The data that supports the findings of this study are available from the corresponding author upon reasonable request.
\section*{Code availability}
The code that supports the findings of this study are available from the corresponding author upon reasonable request.

\section*{Acknowledgments}
We thank Robert Spero for his insightful comments and anonymous referees for their useful feedback.

This research was funded by the Australian Research Council Centre of Excellence CE170100012, Australian Research Council Centre of Excellence CE170100004, Australian Research Council Centre of Excellence CE170100009, Laureate Fellowship FL150100019 and the Australian Government Research Training Program Scholarship.

\section*{Author Contribution}
LC and KMcK evaluated the impact of TDI on satellite geodesy. LC, GG, KMcK, SA and PKL contributed to schemes to remove accelerometer noise. LC, TM and SA performed the 3D simulations. All authors contributed to schemes for eliminating shot noise and writing the paper.
\section*{Competing Interests}
The authors declare no competing interests.

\bibliography{GRACE_bib2.bib}

\begin{thebibliography}{10}
\expandafter\ifx\csname url\endcsname\relax
  \def\url#1{\texttt{#1}}\fi
\expandafter\ifx\csname urlprefix\endcsname\relax\def\urlprefix{URL }\fi
\providecommand{\bibinfo}[2]{#2}
\providecommand{\eprint}[2][]{\url{#2}}

\bibitem{wolff1969direct}
\bibinfo{author}{Wolff, M.}
\newblock \bibinfo{title}{Direct measurements of the earth's gravitational
  potential using a satellite pair}.
\newblock \emph{\bibinfo{journal}{J. Geophys. Res.}}
  \textbf{\bibinfo{volume}{74}}, \bibinfo{pages}{5295--5300}
  (\bibinfo{year}{1969}).

\bibitem{wahr2004time}
\bibinfo{author}{Wahr, J.}, \bibinfo{author}{Swenson, S.},
  \bibinfo{author}{Zlotnicki, V.} \& \bibinfo{author}{Velicogna, I.}
\newblock \bibinfo{title}{Time-variable gravity from grace: First results}.
\newblock \emph{\bibinfo{journal}{Geophys. Res. Lett.}}
  \textbf{\bibinfo{volume}{31}}, \bibinfo{pages}{11} (\bibinfo{year}{2004}).

\bibitem{tapley2004grace}
\bibinfo{author}{Tapley, B.}, \bibinfo{author}{Bettadpur, S.},
  \bibinfo{author}{Ries, J.}, \bibinfo{author}{Thompson, P.} \&
  \bibinfo{author}{Watkins, M.}
\newblock \bibinfo{title}{Grace measurements of mass variability in the earth
  system}.
\newblock \emph{\bibinfo{journal}{Science}} \textbf{\bibinfo{volume}{305}},
  \bibinfo{pages}{503--505} (\bibinfo{year}{2004}).

\bibitem{tapley2004gravity}
\bibinfo{author}{Tapley, B.}, \bibinfo{author}{Bettadpur, S.},
  \bibinfo{author}{Watkins, M.} \& \bibinfo{author}{Reigber, C.}
\newblock \bibinfo{title}{The gravity recovery and climate experiment: Mission
  overview and early results}.
\newblock \emph{\bibinfo{journal}{Geophys. Res. Lett.}}
  \textbf{\bibinfo{volume}{31}}, \bibinfo{pages}{9} (\bibinfo{year}{2004}).

\bibitem{kim2003simulation}
\bibinfo{author}{Kim, J.} \& \bibinfo{author}{Tapley, B.}
\newblock \bibinfo{title}{Simulation of dual one-way ranging measurements}.
\newblock \emph{\bibinfo{journal}{J. Spacecr. Rockets}}
  \textbf{\bibinfo{volume}{40}}, \bibinfo{pages}{419--425}
  (\bibinfo{year}{2003}).

\bibitem{abich2019orbit}
\bibinfo{author}{Abich, K.} \emph{et~al.}
\newblock \bibinfo{title}{In-orbit performance of the grace follow-on laser
  ranging interferometer}.
\newblock \emph{\bibinfo{journal}{Phys. Rev. Lett.}}
  \textbf{\bibinfo{volume}{123}}, \bibinfo{pages}{031101}
  (\bibinfo{year}{2019}).

\bibitem{sheard2012intersatellite}
\bibinfo{author}{Sheard, B.} \emph{et~al.}
\newblock \bibinfo{title}{Intersatellite laser ranging instrument for the grace
  follow-on mission}.
\newblock \emph{\bibinfo{journal}{J Geod.}} \textbf{\bibinfo{volume}{86}},
  \bibinfo{pages}{1083--1095} (\bibinfo{year}{2012}).

\bibitem{amaro2017laser}
\bibinfo{author}{Amaro-Seoane, P.} \emph{et~al.}
\newblock \bibinfo{title}{Laser interferometer space antenna}.
\newblock \emph{\bibinfo{journal}{Preprint at
  \url{https://arxiv.org/abs/1702.00786}}}  (\bibinfo{year}{2017}).

\bibitem{khatri2019spooky}
\bibinfo{author}{Khatri, S.}, \bibinfo{author}{Brady, A.},
  \bibinfo{author}{Desporte, R.}, \bibinfo{author}{Bart, M.} \&
  \bibinfo{author}{Dowling, J.}
\newblock \bibinfo{title}{Spooky action at a global distance $-$ resource-rate
  analysis of a space-based entanglement-distribution network for the quantum
  internet}.
\newblock \emph{\bibinfo{journal}{NPJ Quantum Inf.}}
  \textbf{\bibinfo{volume}{7}}, \bibinfo{pages}{4} (\bibinfo{year}{2021}).

\bibitem{aspelmeyer2003long}
\bibinfo{author}{Aspelmeyer, M.}, \bibinfo{author}{Jennewein, T.},
  \bibinfo{author}{Pfennigbauer, M.}, \bibinfo{author}{Leeb, W.} \&
  \bibinfo{author}{Zeilinger, A.}
\newblock \bibinfo{title}{Long-distance quantum communication with entangled
  photons using satellites}.
\newblock \emph{\bibinfo{journal}{IEEE J. Sel. Top. Quantum Electron.}}
  \textbf{\bibinfo{volume}{9}}, \bibinfo{pages}{1541--1551}
  (\bibinfo{year}{2003}).

\bibitem{simon2017towards}
\bibinfo{author}{Simon, C.}
\newblock \bibinfo{title}{Towards a global quantum network}.
\newblock \emph{\bibinfo{journal}{Nat. Photonics}}
  \textbf{\bibinfo{volume}{11}}, \bibinfo{pages}{678--680}
  (\bibinfo{year}{2017}).

\bibitem{vallone2015experimental}
\bibinfo{author}{Vallone, G.} \emph{et~al.}
\newblock \bibinfo{title}{Experimental satellite quantum communications}.
\newblock \emph{\bibinfo{journal}{Phys. Rev. Lett.}}
  \textbf{\bibinfo{volume}{115}}, \bibinfo{pages}{040502}
  (\bibinfo{year}{2015}).

\bibitem{liao2017satellite}
\bibinfo{author}{Liao, S.} \emph{et~al.}
\newblock \bibinfo{title}{Satellite-to-ground quantum key distribution}.
\newblock \emph{\bibinfo{journal}{Nature}} \textbf{\bibinfo{volume}{549}},
  \bibinfo{pages}{43--47} (\bibinfo{year}{2017}).

\bibitem{liao2018satellite}
\bibinfo{author}{Liao, S.} \emph{et~al.}
\newblock \bibinfo{title}{Satellite-relayed intercontinental quantum network}.
\newblock \emph{\bibinfo{journal}{Phys. Rev. Lett.}}
  \textbf{\bibinfo{volume}{120}}, \bibinfo{pages}{030501}
  (\bibinfo{year}{2018}).

\bibitem{bedington2017progress}
\bibinfo{author}{Bedington, R.}, \bibinfo{author}{Arrazola, J.} \&
  \bibinfo{author}{Ling, A.}
\newblock \bibinfo{title}{Progress in satellite quantum key distribution}.
\newblock \emph{\bibinfo{journal}{NPJ Quantum Inf.}}
  \textbf{\bibinfo{volume}{3}}, \bibinfo{pages}{1--13} (\bibinfo{year}{2017}).

\bibitem{caves1981quantum}
\bibinfo{author}{Caves, C.}
\newblock \bibinfo{title}{Quantum-mechanical noise in an interferometer}.
\newblock \emph{\bibinfo{journal}{Phys. Rev. D}} \textbf{\bibinfo{volume}{23}},
  \bibinfo{pages}{1693} (\bibinfo{year}{1981}).

\bibitem{christophe2015new}
\bibinfo{author}{Christophe, B.} \emph{et~al.}
\newblock \bibinfo{title}{A new generation of ultra-sensitive electrostatic
  accelerometers for grace follow-on and towards the next generation gravity
  missions}.
\newblock \emph{\bibinfo{journal}{Acta Astronaut.}}
  \textbf{\bibinfo{volume}{117}}, \bibinfo{pages}{1--7} (\bibinfo{year}{2015}).

\bibitem{dobslaw2016modeling}
\bibinfo{author}{Dobslaw, H.} \emph{et~al.}
\newblock \bibinfo{title}{Modeling of present-day atmosphere and ocean
  non-tidal de-aliasing errors for future gravity mission simulations}.
\newblock \emph{\bibinfo{journal}{J. Geod.}} \textbf{\bibinfo{volume}{90}},
  \bibinfo{pages}{423--436} (\bibinfo{year}{2016}).

\bibitem{wegener2020tilt}
\bibinfo{author}{Wegener, H.}, \bibinfo{author}{M{\"u}ller, V.},
  \bibinfo{author}{Heinzel, G.} \& \bibinfo{author}{Misfeldt, M.}
\newblock \bibinfo{title}{Tilt-to-length coupling in the grace follow-on laser
  ranging interferometer}.
\newblock \emph{\bibinfo{journal}{J. Spacecr. Rockets}}
  \textbf{\bibinfo{volume}{57}}, \bibinfo{pages}{1362--1372}
  (\bibinfo{year}{2020}).

\bibitem{tinto1999cancellation}
\bibinfo{author}{Tinto, M.} \& \bibinfo{author}{Armstrong, J.}
\newblock \bibinfo{title}{Cancellation of laser noise in an unequal-arm
  interferometer detector of gravitational radiation}.
\newblock \emph{\bibinfo{journal}{Phys. Rev. D}} \textbf{\bibinfo{volume}{59}},
  \bibinfo{pages}{102003} (\bibinfo{year}{1999}).

\bibitem{armstrong1999time}
\bibinfo{author}{Armstrong, J.}, \bibinfo{author}{Estabrook, F.} \&
  \bibinfo{author}{Tinto, M.}
\newblock \bibinfo{title}{Time-delay interferometry for space-based
  gravitational wave searches}.
\newblock \emph{\bibinfo{journal}{Astrophys. J.}}
  \textbf{\bibinfo{volume}{527}}, \bibinfo{pages}{814} (\bibinfo{year}{1999}).

\bibitem{tinto2002time}
\bibinfo{author}{Tinto, M.}, \bibinfo{author}{Estabrook, F.} \&
  \bibinfo{author}{Armstrong, J.}
\newblock \bibinfo{title}{Time-delay interferometry for lisa}.
\newblock \emph{\bibinfo{journal}{Phys. Rev. D}} \textbf{\bibinfo{volume}{65}},
  \bibinfo{pages}{082003} (\bibinfo{year}{2002}).

\bibitem{tinto2003implementation}
\bibinfo{author}{Tinto, M.}, \bibinfo{author}{Shaddock, D.},
  \bibinfo{author}{Sylvestre, J.} \& \bibinfo{author}{Armstrong, J.}
\newblock \bibinfo{title}{Implementation of time-delay interferometry for
  lisa}.
\newblock \emph{\bibinfo{journal}{Phys. Rev. D}} \textbf{\bibinfo{volume}{67}},
  \bibinfo{pages}{122003} (\bibinfo{year}{2003}).

\bibitem{tinto2004time}
\bibinfo{author}{Tinto, M.}, \bibinfo{author}{Estabrook, F.} \&
  \bibinfo{author}{Armstrong, J.}
\newblock \bibinfo{title}{Time delay interferometry with moving spacecraft
  arrays}.
\newblock \emph{\bibinfo{journal}{Phys. Rev. D}} \textbf{\bibinfo{volume}{69}},
  \bibinfo{pages}{082001} (\bibinfo{year}{2004}).

\bibitem{francis2015tone}
\bibinfo{author}{Francis, S.} \emph{et~al.}
\newblock \bibinfo{title}{Tone-assisted time delay interferometry on grace
  follow-on}.
\newblock \emph{\bibinfo{journal}{Phys. Rev. D}} \textbf{\bibinfo{volume}{92}},
  \bibinfo{pages}{012005} (\bibinfo{year}{2015}).

\bibitem{sneeuw2005satellite}
\bibinfo{author}{Sneeuw, N.} \& \bibinfo{author}{Schaub, H.}
\newblock \bibinfo{title}{Satellite clusters for future gravity field
  missions}.
\newblock In \emph{\bibinfo{booktitle}{Gravity, geoid and space missions}},
  \bibinfo{pages}{12--17} (\bibinfo{publisher}{Springer, Berlin},
  \bibinfo{year}{2005}).

\bibitem{sharifi2007gravity}
\bibinfo{author}{Sharifi, M.}, \bibinfo{author}{Sneeuw, N.} \&
  \bibinfo{author}{Keller, W.}
\newblock \bibinfo{title}{Gravity recovery capability of four generic satellite
  formations}.
\newblock \emph{\bibinfo{journal}{Gravity Field of the Earth. General Command
  of Mapping, ISSN. \textit{Special issue}}} \textbf{\bibinfo{volume}{18}},
  \bibinfo{pages}{211--216} (\bibinfo{year}{2007}).

\bibitem{reubelt2010quick}
\bibinfo{author}{Reubelt, T.}, \bibinfo{author}{Sneeuw, N.} \&
  \bibinfo{author}{Iran-Pour, S.}
\newblock \bibinfo{title}{Quick-look gravity field analysis of formation
  scenarios selection}.
\newblock \emph{\bibinfo{journal}{GEOTECHNOLOGIEN Science Report}}
  \textbf{\bibinfo{volume}{17}}, \bibinfo{pages}{126--133}
  (\bibinfo{year}{2010}).

\bibitem{savcenko2012eot11a}
\bibinfo{author}{Savcenko, R.} \& \bibinfo{author}{Bosch, W.}
\newblock \bibinfo{title}{Eot11a-empirical ocean tide model from multi-mission
  satellite altimetry}.
\newblock \emph{\bibinfo{journal}{DGFI Report No. 89}}  (\bibinfo{year}{2012}).

\bibitem{kvas2019grace}
\bibinfo{author}{Kvas, A.} \& \bibinfo{author}{Mayer-G{\"u}rr, T.}
\newblock \bibinfo{title}{Grace gravity field recovery with background model
  uncertainties}.
\newblock \emph{\bibinfo{journal}{J. Geod.}} \textbf{\bibinfo{volume}{93}},
  \bibinfo{pages}{2543--2552} (\bibinfo{year}{2019}).

\bibitem{spero2021point}
\bibinfo{author}{Spero, R.}
\newblock \bibinfo{title}{Point-mass sensitivity of gravimetric satellites}.
\newblock \emph{\bibinfo{journal}{Adv. Space Res.}}
  \textbf{\bibinfo{volume}{67}}, \bibinfo{pages}{1656--1664}
  (\bibinfo{year}{2021}).

\bibitem{touboul1999electrostatic}
\bibinfo{author}{Touboul, P.}, \bibinfo{author}{Foulon, B.} \&
  \bibinfo{author}{Willemenot, E.}
\newblock \bibinfo{title}{Electrostatic space accelerometers for present and
  future missions}.
\newblock \emph{\bibinfo{journal}{Acta Astronaut.}}
  \textbf{\bibinfo{volume}{45}}, \bibinfo{pages}{605--617}
  (\bibinfo{year}{1999}).

\bibitem{estabrook2003syzygy}
\bibinfo{author}{Estabrook, F.}, \bibinfo{author}{Armstrong, J.},
  \bibinfo{author}{Tinto, M.} \& \bibinfo{author}{Folkner, W.}
\newblock \bibinfo{title}{Syzygy: A straight interferometric spacecraft system
  for gravity wave observations}.
\newblock \emph{\bibinfo{journal}{Phys. Rev. D}} \textbf{\bibinfo{volume}{68}},
  \bibinfo{pages}{062001} (\bibinfo{year}{2003}).

\bibitem{kroes2005precise}
\bibinfo{author}{Kroes, R.}, \bibinfo{author}{Montenbruck, O.},
  \bibinfo{author}{Bertiger, W.} \& \bibinfo{author}{Visser, P.}
\newblock \bibinfo{title}{Precise grace baseline determination using gps}.
\newblock \emph{\bibinfo{journal}{GPS Solut.}} \textbf{\bibinfo{volume}{9}},
  \bibinfo{pages}{21--31} (\bibinfo{year}{2005}).

\bibitem{wu2006real}
\bibinfo{author}{Wu, S.} \& \bibinfo{author}{Bar-Sever, Y.}
\newblock \bibinfo{title}{Real-time sub-cm differential orbit determination of
  two low-earth orbiters with gps bias fixing}.
\newblock \emph{\bibinfo{journal}{JPL Technical Reports Server}}
  (\bibinfo{year}{2006}).

\bibitem{schnabel2010quantum}
\bibinfo{author}{Schnabel, R.}, \bibinfo{author}{Mavalvala, N.},
  \bibinfo{author}{McClelland, D.} \& \bibinfo{author}{Lam, P.}
\newblock \bibinfo{title}{Quantum metrology for gravitational wave astronomy}.
\newblock \emph{\bibinfo{journal}{Nat. Commun.}} \textbf{\bibinfo{volume}{1}},
  \bibinfo{pages}{1--10} (\bibinfo{year}{2010}).

\bibitem{armano2018beyond}
\bibinfo{author}{Armano, M.} \emph{et~al.}
\newblock \bibinfo{title}{Beyond the required lisa free-fall performance: new
  lisa pathfinder results down to 20 $\mu$ hz}.
\newblock \emph{\bibinfo{journal}{Phys. Rev. Lett.}}
  \textbf{\bibinfo{volume}{120}}, \bibinfo{pages}{061101}
  (\bibinfo{year}{2018}).

\bibitem{christophe2010orbit}
\bibinfo{author}{Christophe, B.}, \bibinfo{author}{Marque, J.} \&
  \bibinfo{author}{Foulon, B.}
\newblock \bibinfo{title}{In-orbit data verification of the accelerometers of
  the esa goce mission}.
\newblock In \emph{\bibinfo{booktitle}{SF2A-2010: Proceedings of the Annual
  meeting of the French Society of Astronomy and Astrophysics}},
  vol.~\bibinfo{volume}{1}, \bibinfo{pages}{113} (\bibinfo{year}{Marseille,
  2010}).

\bibitem{alvarez2021simplified}
\bibinfo{author}{Alvarez, A.~D.} \emph{et~al.}
\newblock \bibinfo{title}{A simplified gravitational reference sensor for
  satellite geodesy}.
\newblock \emph{\bibinfo{journal}{Preprint at
  \url{https://arxiv.org/abs/2107.08545}}}  (\bibinfo{year}{2021}).

\bibitem{bandikova2019grace}
\bibinfo{author}{Bandikova, T.}, \bibinfo{author}{McCullough, C.},
  \bibinfo{author}{Kruizinga, G.}, \bibinfo{author}{Save, H.} \&
  \bibinfo{author}{Christophe, B.}
\newblock \bibinfo{title}{Grace accelerometer data transplant}.
\newblock \emph{\bibinfo{journal}{Adv. Space Res.}}
  \textbf{\bibinfo{volume}{64}}, \bibinfo{pages}{623--644}
  (\bibinfo{year}{2019}).

\bibitem{demkowicz2013fundamental}
\bibinfo{author}{Demkowicz-Dobrza{\'n}ski, R.}, \bibinfo{author}{Banaszek, K.}
  \& \bibinfo{author}{Schnabel, R.}
\newblock \bibinfo{title}{Fundamental quantum interferometry bound for the
  squeezed-light-enhanced gravitational wave detector geo 600}.
\newblock \emph{\bibinfo{journal}{Phys. Rev. A}} \textbf{\bibinfo{volume}{88}},
  \bibinfo{pages}{041802} (\bibinfo{year}{2013}).

\bibitem{berni2015ab}
\bibinfo{author}{Berni, A.} \emph{et~al.}
\newblock \bibinfo{title}{Ab initio quantum-enhanced optical phase estimation
  using real-time feedback control}.
\newblock \emph{\bibinfo{journal}{Nat. Photonics}}
  \textbf{\bibinfo{volume}{9}}, \bibinfo{pages}{577--581}
  (\bibinfo{year}{2015}).

\bibitem{goda2008quantum}
\bibinfo{author}{Goda, K.} \emph{et~al.}
\newblock \bibinfo{title}{A quantum-enhanced prototype gravitational-wave
  detector}.
\newblock \emph{\bibinfo{journal}{Nat. Phys.}} \textbf{\bibinfo{volume}{4}},
  \bibinfo{pages}{472--476} (\bibinfo{year}{2008}).

\bibitem{xiao1987precision}
\bibinfo{author}{Xiao, M.}, \bibinfo{author}{Wu, L.} \&
  \bibinfo{author}{Kimble, H.}
\newblock \bibinfo{title}{Precision measurement beyond the shot-noise limit}.
\newblock \emph{\bibinfo{journal}{Phys. Rev. Lett.}}
  \textbf{\bibinfo{volume}{59}}, \bibinfo{pages}{278} (\bibinfo{year}{1987}).

\bibitem{gladstone2016atmosphere}
\bibinfo{author}{Gladstone, G.} \emph{et~al.}
\newblock \bibinfo{title}{The atmosphere of pluto as observed by new horizons}.
\newblock \emph{\bibinfo{journal}{Science}} \textbf{\bibinfo{volume}{351}},
  \bibinfo{pages}{6279} (\bibinfo{year}{2016}).

\bibitem{aasi2013enhanced}
\bibinfo{author}{Aasi, J.} \emph{et~al.}
\newblock \bibinfo{title}{Enhanced sensitivity of the ligo gravitational wave
  detector by using squeezed states of light}.
\newblock \emph{\bibinfo{journal}{Nat. Photonics}}
  \textbf{\bibinfo{volume}{7}}, \bibinfo{pages}{613--619}
  (\bibinfo{year}{2013}).

\bibitem{khwaja2019low}
\bibinfo{author}{Khwaja, T.} \& \bibinfo{author}{Reza, S.}
\newblock \bibinfo{title}{Low-cost gaussian beam profiling with circular irises
  and apertures}.
\newblock \emph{\bibinfo{journal}{Appl. Opt.}} \textbf{\bibinfo{volume}{58}},
  \bibinfo{pages}{1048--1056} (\bibinfo{year}{2019}).

\bibitem{tinto2014time}
\bibinfo{author}{Tinto, M.} \& \bibinfo{author}{Dhurandhar, S.}
\newblock \bibinfo{title}{Time-delay interferometry}.
\newblock \emph{\bibinfo{journal}{Living Rev. Relativ.}}
  \textbf{\bibinfo{volume}{17}}, \bibinfo{pages}{6} (\bibinfo{year}{2014}).

\bibitem{flanagan1998measuring}
\bibinfo{author}{Flanagan, E.} \& \bibinfo{author}{Hughes, S.}
\newblock \bibinfo{title}{Measuring gravitational waves from binary black hole
  coalescences. i. signal to noise for inspiral, merger, and ringdown}.
\newblock \emph{\bibinfo{journal}{Phys. Rev. D}} \textbf{\bibinfo{volume}{57}},
  \bibinfo{pages}{4535} (\bibinfo{year}{1998}).

\bibitem{dorner2009optimal}
\bibinfo{author}{Dorner, U.} \emph{et~al.}
\newblock \bibinfo{title}{Optimal quantum phase estimation}.
\newblock \emph{\bibinfo{journal}{Phys. Rev. Lett.}}
  \textbf{\bibinfo{volume}{102}}, \bibinfo{pages}{040403}
  (\bibinfo{year}{2009}).

\bibitem{armano2016sub}
\bibinfo{author}{Armano, M.} \emph{et~al.}
\newblock \bibinfo{title}{Sub-femto-g free fall for space-based gravitational
  wave observatories: Lisa pathfinder results}.
\newblock \emph{\bibinfo{journal}{Phys. Rev. Lett.}}
  \textbf{\bibinfo{volume}{116}}, \bibinfo{pages}{231101}
  (\bibinfo{year}{2016}).

\end{thebibliography}
\bibliographystyle{naturemag}

\appendix

\section{Fourier domain two satellite acceleration range signal}
\label{apen:sigderiv}
We consider a satellite travelling with an initial velocity  $v_0$ at a height $h$ above a plane. There is a point mass $M$ on the plane, which at time, $t=0$, is a horizontal distance $x$ from the satellite, so that the total distance from the satellite to the mass is $\sqrt{h^2+x^2}$. Assuming that the along track position is not affected significantly by the gravitational attraction, $x=v_0t$, the acceleration in the along track direction for a single satellite in the time domain is given by
\begin{equation}
a(t)=-\frac{GMtv_0}{(h^2+t^2v_0^2)^{3/2}}\;,
\end{equation}
where $G=6.67\times10^{-11}\text{ m}^3\text{kg}^{-1}\text{s}^{-2}$ is the gravitational constant. Converting to the frequency domain gives
\begin{equation}
a(f)=-\frac{GMv_0}{h^3}\int_{-\infty}^{\infty}\frac{e^{-2\pi \mathrm{i} f t}t}{(1+\frac{t^2v_0^2}{h^2})^{3/2}}dt\;.
\end{equation}
Using the substitution $u=tv_0/h$, this can be written
\begin{equation}
a(f)=\frac{-GM}{hv_0}\int_{-\infty}^{\infty}\frac{e^{-2\pi \mathrm{i} f\frac{h}{v_0} u} u}{(1+u^2)^{3/2}}du\;.
\end{equation}
Through the product rule (using $b=e^{-2\pi \mathrm{i} f u h/v_0 }, dc=u/(1+u^2)^{3/2}$) this becomes 
\begin{equation}
\begin{split}
a(f)=&\frac{-GM}{h v_0}\biggl[ \frac{-e^{-2\pi \mathrm{i} f \frac{h}{v_0}u} }{(1+u^2)^{1/2}}\Bigg\rvert_{u=-\infty}^{u=\infty} \\
&-\int_{-\infty}^{\infty}\frac{2\pi \mathrm{i} f \frac{h}{v_0}e^{-2\pi \mathrm{i} f  \frac{h}{v_0}u}}{\sqrt{1+u^2}}du\biggr]\;.
\end{split}
\end{equation}
The first term is zero and the second term can be identified as a multiple of the modified Bessel function of the second kind, order 0, $K_0$, giving
\begin{equation}
\label{acc_0}
a(f)=\frac{4 \pi \mathrm{i} f G M}{v_0^2}K_0\left(\frac{2\pi f}{f_h}\right)\;,
\end{equation}
where $f_h=v_0/h$. This is the acceleration of a single satellite in the frequency domain. For satellite geodesy missions we are interested in how the range between a pair of satellites changes in time and so we consider the range acceleration. %To obtain 

The range acceleration is the differential acceleration of the two satellites, obtained by subtracting one signal from the other. The effect of subtracting one signal from another is to multiply the signal by a sin term. 
\begin{equation}
\begin{split}
S(t)&=a(t)-a(t-\tau)\\
\mathcal{S}(f)&=\mathcal{F}(S(t))=a(f)(1-e^{-2\pi \mathrm{i} f \tau})\\
\abs{\mathcal{S}(f)}&=\abs{a(f)}\abs{e^{-\pi \mathrm{i} f \tau}}\abs{(e^{\pi \mathrm{i} f \tau}-e^{-\pi \mathrm{i} f \tau})} \\
\abs{\mathcal{S}(f)}&=\abs{a(f)}\abs{2\text{ sin}(\pi f \tau)}\;.
\end{split}
\end{equation}
For a GRACE-like mission one satellite follows another along a very similar trajectory, corresponding to a delay of $\tau_S=L/v_0$, where $L$ is the satellite separation and the subscript $S$ denotes that this time of flight corresponds to the satellite velocity. Thus for a single satellite pair the range acceleration in the frequency domain is
\begin{equation}
\label{eq:arsig}
\abs{a_R(f)}=\frac{8 \pi f G M}{v_0^2}K_0\left(\frac{2\pi f}{f_h}\right)\abs{\text{sin}\left(\frac{2\pi f}{f_L}\right)}\;,
\end{equation}
where $f_L=2v_0/L$. As the current GRACE-FO mission measures twice the phase shift between the two satellites we scale this expression by a factor of 2 in Eq.~\eqref{eq:acc_twosat} in the main text. There are nulls in the signal at certain frequencies, corresponding to $f=\text{n}v_0/L$, where n is any integer. These nulls can be seen in Fig.~\ref{fig:noise_spec}. Intuitively any signal with a period equal to the satellite separation time, $\tau_S$, will not be observed in the ranging signal as it will affect both satellites in the same manner. The same is true for any period which is an integer fraction of $\tau_S$. This model was presented in its entirety in Ref.~\cite{spero2021point}.

\section{Laser phase noise spectrum}
\label{lpnScaleLin}
We note here why the laser phase noise scales linearly with distance. From Eq.~\eqref{eq:maintxtlpn} in the main text, the laser phase noise in the current GRACE-FO measurement is
\begin{equation}
C_\text{tot}(t)=C_1(t)-C_1(t-2\tau_{12})\;,
\end{equation}
where $C_i(t)$ denotes the phase noise of the laser on satellite $i$ at time $t$ and $\tau_{12}=L/c$ denotes the single trip time of flight for light between the two satellites, where $c$ is the speed of light. Note that $\tau_{12}$ is different to $\tau_S$. Taking the Fourier transform of this expression we obtain 
\begin{equation}
C_\text{tot}(f)=C_1(f)(1-e^{-2\pi \mathrm{i} f 2\tau_{12}})\;.
\end{equation}
As the $\tau_{12}$ term refers to a time of flight for light, this will be very small and so using the small angle approximation the absolute value becomes
\begin{equation}
\abs{C_\text{tot}(f)}\approx\abs{C_1(f)}\abs{4\pi  f \tau_{12}}\;.
\end{equation}
We see that the laser phase noise spectrum scales linearly with $\tau_{12}$ or equivalently with $L$, the distance between the satellites. 

\section{Quantum noise spectrum}
\label{sec:loss}
\begin{figure*}[t]
\centering
\includegraphics[width=\textwidth]{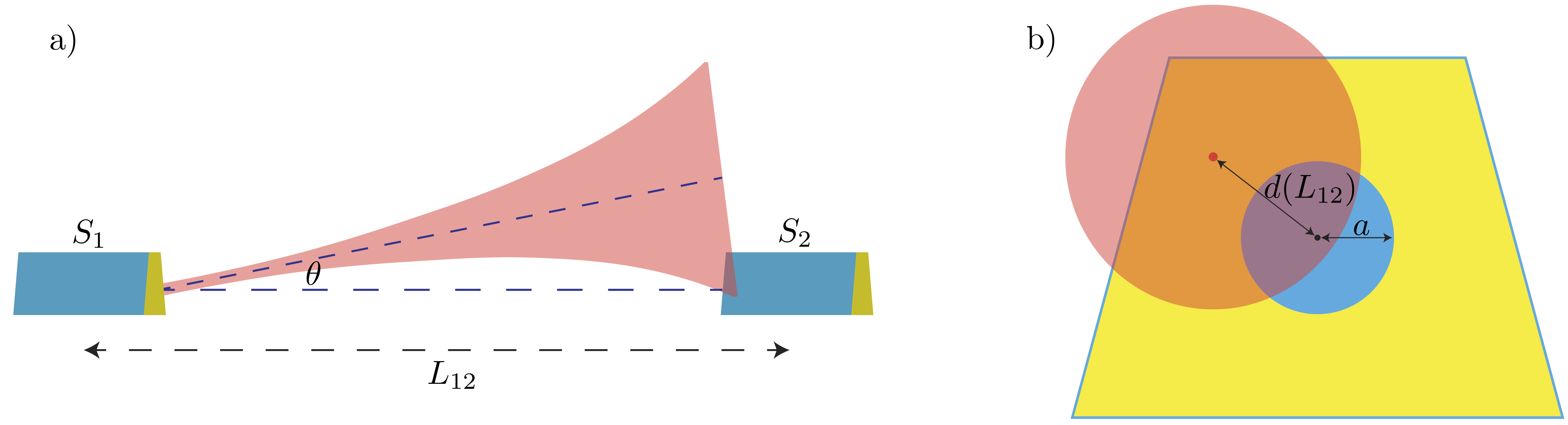}
\caption{\textbf{Optical loss mechanism.} a) Through beam diffraction and imperfect beam alignment, most of the power leaving the first satellite does not arrive at the second satellite. b) The receiving aperture on the second satellite has a radius $a$ and the arriving beam is off centre by a distance $d(L)$. As the satellite separation $L$ becomes larger the beam is further off centre for a fixed pointing angle error. }
\label{fig:sn_mech}
\end{figure*}
The are two major sources of optical loss in satellite-to-satellite communications. These are beam diffraction and beam misalignment (pointing error), which ensure that not all of the optical power leaving the first satellite reaches the second satellite. The amount of optical loss is governed by the properties of Gaussian beams. A Gaussian beam expands as it propagates meaning the incoming beam waist at the second satellite is considerably larger than the receiving optics and so most of the light is lost. The amount of optical loss is therefore a function of satellite separation, $L$. For a Gaussian beam the beam radius at a distance $L$ is 
\begin{equation}
w(L)=w_{0}\sqrt{1+\left(\frac{L}{z_{R}}\right)^{2}}\;,
\end{equation}
where $w_{0}$ is the initial beam waist and
\begin{equation}
z_{R}=\frac{\pi w_{0}^{2}}{\lambda}\;,
\end{equation}
is the Rayleigh range, where $\lambda$ is the wavelength of the light. We assume $\lambda=1064$ nm throughout as this is the wavelength of the laser ranging interferometer in the current mission~\cite{abich2019orbit}. It can then be calculated that the power passing through an aperture of radius $a$, at a distance $L$ is given by
\begin{equation}
P(a, L)=P_{0}[1-e^{\frac{-2a^{2}}{w(L)^{2}}}]\;,
\end{equation}
where $P_0$ is the initial power. This can be modelled as a lossy quantum channel with transmissivity $\eta$, such that $P(a, L)=\eta P_{0}$. Additionally if the beam is not centred (i.e. the centre of the Gaussian beam does not go directly through the centre of the aperture), it is known that this corresponds to a transmissivity of \cite{khwaja2019low}
\begin{equation}
\begin{split}
\eta(a,L)=\text{ }&\text{exp}(\frac{-2d(L)^{2}}{w(L)^{2}})  \sum_{k=0}^{\infty}\biggl(\frac{2^kd(L)^{2k}}{w(L)^{2k}k!}  \\
&\left(1-\text{exp}\left(-\frac{2a^2}{w(L)^2}\right)\sum_{i=0}^{k}\frac{2^ia^{2i}}{w(L)^{2i}i!} \right)\biggr)\;,
\end{split}
\end{equation}
where $d(L)$ is the distance off centre, i.e. the distance between the centre of the Gaussian beam and the centre of the aperture and $a$ is the radius of the aperture (a few cm on GRACE-FO). $d(L)$ is function of satellite separation $L$ because the system will have a certain angular resolution (mRad for GRACE-FO) which gets worse with larger distances. For small offsets, with an angular resolution $\theta$, the distance off centre is
\begin{equation}
d(L)=L\text{tan}(\theta)\approx L \theta\;.
\end{equation}
This loss mechanism is shown in Fig.~\ref{fig:sn_mech}.

Thus for a given set of satellite parameters every distance corresponds to a certain optical loss, which can be modelled as a beamsplitter of transmissivity $\eta$. We assume the input to the second arm of the beam-splitter is vacuum as there are very few thermal photons at optical frequencies in space. With this approach it is simple to examine the impact this channel will have on the mean and covariance matrix of a Gaussian state. After travelling from one satellite to another, a coherent state, initially $\hat{D}(\alpha)\ket{0}=\ket{\alpha}$, where $\hat{D}(\alpha)$ is the displacement operator, will have mean and covariance defined by
\begin{align}
\langle \hat{\alpha} \rangle &=
\begin{bmatrix}
2\sqrt{\eta}\alpha\\
0\\
\end{bmatrix}\;,\\
\Sigma_{\alpha}&=I\;.
\end{align}
where I is the 2x2 identity matrix. The average photon number for this coherent state is now $\langle \hat{n} \rangle =\abs{\eta\alpha^{2}}$.  A displaced squeezed state, $\hat{D}(\alpha)\hat{\mathcal{E}}(r)\ket{0}$, where $\hat{\mathcal{E}}(r)$ is the squeezing operator and r is the initial squeezing parameter, will suffer a degradation of both the amount and the purity of squeezing as the satellite separation is increased. The effect of the beam-splitter on the covariance matrix of a squeezed beam is given by:
\begin{equation}
\Sigma_{\alpha_{r}}=
\begin{bmatrix}
\eta e^{-2r}+(1-\eta)&0\\
0&\eta e^{2r}+(1-\eta)\\
\end{bmatrix}\;.
\end{equation}
This approach offers a simple way to determine how the squeezing level and mean photon number are affected by satellite separation. Hence given a certain input state and satellite separation, the received state is completely determined. From this the quantum noise level is determined. Given an initial power $P_0$, the power received at the second satellite is $P_{\text{rec}}=\eta(L)P_0$. The number of photons received per second is then $\alpha^2=P_{\text{rec}}/(hf)$, where $h$ is Plancks constant and $f$ is the frequency of the light. The minimum allowed standard deviation for measuring phase when quantum noise limited is
\begin{equation}
\triangle\phi=\frac{\sqrt{e^{-2r_{\mathrm{eff}}}}}{\sqrt{4\eta\alpha^2}}\;,
\label{eq:SNspec}
\end{equation}
where $e^{-2r_{\mathrm{eff}}}=\eta e^{-2r}+(1-\eta)$ defines $r_{\mathrm{eff}}$, the effective squeezing parameter the second satellite receives. Eq.~\eqref{eq:SNspec} assumes that homodyne detection has been employed to keep the measurement locked to the squeezed quadrature and does not account for optical inefficiencies on board the receiving satellite. The ranging uncertainty is the phase uncertainty multiplied by $\lambda/(2\pi)$, where $\lambda$ is the wavelength of the light. Quantum noise corresponds to a white noise spectrum and can be converted to an acceleration noise spectrum by multiplying by $(2\pi f)^2$.

\section{Alternative laser phase noise free combinations}
\label{apen:altlpnfree}
In the main text only one TDI combination was explicitly considered, that with a single laser at the middle satellite, formation $\alpha^T_3$. This laser light is split and directed towards the two outer satellites, where it is reflected back towards the middle satellite. We now show that many other combinations are possible. For the single laser combinations it is possible to have the laser on either the first or third satellites. This light is sent to the satellites without a laser and reflected back to the satellite with the laser. This arrangement may lead to the interaction of the light with itself as both paths overlap which can lead to further complications. This is avoidable however, for example by frequency shifting the light at the distant satellite. With the laser on the first satellite the following combination is laser phase noise free
\begin{equation}
\begin{split}
&\text{ }[2\hat{\phi}_{12}(t)-2\hat{\phi}_{13}(t)-2\hat{\phi}_{12}(t-2\tau_{12}-2\tau_{23})\\
&+2\hat{\phi}_{13}(t-2\tau_{12})] \\ 
=&\text{ }[2\phi_{12}(t)-2\phi_{13}(t)-(2\phi_{12}(t-2\tau_{12}-2\tau_{23})\\
&-2\phi_{13}(t-2\tau_{12}))]\\
&+(QN_{12}(t)-QN_{12}(t-2\tau_{12}-2\tau_{23}))\\
&+(QN_{13}(t-2\tau_{12})-QN_{13}(t))\\
&+(AN_{12}(t)-AN_{12}(t-2\tau_{12}-2\tau_{23}))\\
&+(AN_{13}(t-2\tau_{12})-AN_{13}(t))\;.
\end{split}
\end{equation}
In this case we can assume the two accelerometer noise terms have the same frequency spectrum but we cannot assume the two quantum noise terms have the same spectrum as they correspond to light which has travelled different distances and hence correspond to different quantum noise levels. For this configuration the signal is transformed as
\begin{equation}
\begin{split}
\abs{a_{R(TDI)}(f)}=&\frac{8\pi f GM}{v_0^2}\abs{K_0\left( \frac{2\pi f}{f_h} \right)}\\
& \biggl|\biggl(-e^{-2\pi \mathrm{i} f \tau_{S}}+e^{-2\pi \mathrm{i} f (2\tau_{S})}\\
&-e^{-2\pi \mathrm{i} f (4\tau)}+e^{-2\pi \mathrm{i} f (\tau_{S}+4\tau)}\\
&+e^{-2\pi \mathrm{i} f (2\tau)}-e^{-2\pi \mathrm{i} f (2\tau_{S}+2\tau)}\biggr)\biggr|\;,
\end{split}
\end{equation}
where $\tau=\tau_{12}=\tau_{23}$ is the single trip time of flight for light between the two satellites, assuming the satellites separations are equal. Although it is not obvious, in the frequency domain, this is very similar to the combination with the laser at the middle satellite. We consider the accelerometer noise as two separate contributions each with the same noise spectrum but with a different delay, which changes the total accelerometer noise spectrum to $\abs{AN_{13}(f)(1-e^{-2\pi \mathrm{i} f2\tau})-AN_{12}(f)(1-e^{-2\pi \mathrm{i} f4\tau})}$. Similarly the quantum noise spectrum will transform to $\abs{QN_{13}(f)(1-e^{-2\pi \mathrm{i} f2\tau})-QN_{12}(f)(1-e^{-2\pi \mathrm{i} f4\tau})}$. Essentially the same analysis holds for the combination which has the laser on the third satellite. It may be possible to simultaneously operate all three single laser configurations at different wavelengths. This would provide extra information and still allow the removal of ionospheric effects~\cite{kim2003simulation}.

We now consider alternate, multi-laser schemes which have all been explored for LISA~\cite{armstrong1999time}. We examine a 3 laser, 6 measurement configuration which includes acceleration and quantum noise. The 6 measured phases are
\begin{equation}
\hat{\phi}_{12}=\phi_{12}+C_{2}(t-\tau_{12})-C_{1}(t)+AN_{12}(t)+QN_{12}(t) 
\end{equation}
\begin{equation}
\hat{\phi}_{13}=\phi_{13}+C_{3}(t-\tau_{13})-C_{1}(t)+AN_{13}(t)+QN_{13}(t)
\end{equation}
\begin{equation}
\hat{\phi}_{21}=\phi_{12}+C_{1}(t-\tau_{12})-C_{2}(t)+AN_{21}(t)+QN_{21}(t) 
\end{equation}
\begin{equation}
\hat{\phi}_{23}=\phi_{23}+C_{3}(t-\tau_{23})-C_{2}(t)+AN_{23}(t)+QN_{23}(t) 
\end{equation}
\begin{equation}
\hat{\phi}_{31}=\phi_{13}+C_{1}(t-\tau_{13})-C_{3}(t)+AN_{31}(t)+QN_{31}(t)
\end{equation}
\begin{equation}
\hat{\phi}_{32}=\phi_{23}+C_{2}(t-\tau_{23})-C_{3}(t)+AN_{32}(t)+QN_{32}(t)
\end{equation}
where we have assumed the time of flights are symmetric, i.e. $\tau_{12}=\tau_{21}$. From these 6 measurements the following 3 laser phase noise free combinations can be constructed.
\begin{equation}
\begin{split}
\alpha(t)=&\text{ }\hat{\phi}_{13}(t)-\hat{\phi}_{12}(t)+\hat{\phi}_{32}(t-\tau_{13})-\hat{\phi}_{23}(t-\tau_{12})\\
&+\hat{\phi}_{21}(t-\tau_{23}-\tau_{13})-\hat{\phi}_{31}(t-\tau_{23}-\tau_{12}) \\
\end{split}
\end{equation}
\begin{equation}
\begin{split}
\beta(t)=&\text{ }\hat{\phi}_{21}(t)-\hat{\phi}_{23}(t)+\hat{\phi}_{13}(t-\tau_{12})-\hat{\phi}_{31}(t-\tau_{23})\\
&+\hat{\phi}_{32}(t-\tau_{13}-\tau_{12})-\hat{\phi}_{12}(t-\tau_{13}-\tau_{23})\\
\end{split}
\end{equation}
\begin{equation}
\begin{split}
\gamma(t)=&\text{ }\hat{\phi}_{32}(t)-\hat{\phi}_{31}(t)+\hat{\phi}_{21}(t-\tau_{23})-\hat{\phi}_{12}(t-\tau_{13})\\
&+\hat{\phi}_{13}(t-\tau_{23}-\tau_{12})-\hat{\phi}_{23}(t-\tau_{13}-\tau_{12})
\end{split}
\end{equation}
These schemes can be compared to the single laser TDI schemes by examining how the signal and various noise sources are transformed by these combinations. In each combination the signal remaining is
\begin{equation}
\begin{split}
\alpha_{sig}(t)=&\text{ }\phi_{23}(t)-\phi_{23}(t-\tau_{12})\\
&+ \phi_{12}(t-\tau_{12}-2\tau_{23})-\phi_{12}(t-\tau_{12}-\tau_{23}),\\
\end{split}
\end{equation}
\begin{equation}
\begin{split}
\beta_{sig}(t)\approx &\text{ }\phi_{12}(t)-\phi_{23}(t)\\
&-\phi_{12}(t-\tau_{12}-2\tau_{23})+\phi_{23}(t-\tau_{12}-\tau_{23}), \\
\end{split}
\end{equation}
\begin{equation}
\begin{split}
\gamma_{sig}(t)=&\text{ }-\phi_{12}(t)+\phi_{12}(t-\tau_{23})\\
&+\phi_{23}(t-\tau_{12}-\tau_{23})-\phi_{23}(t-2\tau_{12}-\tau_{23}) ,
\end{split}
\end{equation}
where we have used the fact that $\tau_{13}=\tau_{12}+\tau_{23}$ and $\phi_{13}=\phi_{12}+\phi_{23}$. It is possible to convert to the frequency domain, where we see that these combinations result in a greatly reduced signal. We can also note that in each combination no measurement is used twice, and so each combination contains the quantum noise 6 times. Thus the quantum noise is not reduced in the same way that the signal and accelerometer noise are. These schemes do not perform as well as the single laser schemes. This is unfortunate given that two laser schemes are easier to implement. However, the single laser schemes can be converted to multi-laser schemes which do not involve any reflections through phase locking.
\section{Gravitational signal after TDI}
\label{apen:gravsigtdi}
As discussed in the main text, by switching to a three satellite configuration and using TDI, the laser phase noise can be completely removed from the measurement. With a single laser on the middle satellite, sent to the outer satellites and back, formation $\alpha^T_3$, the following combination of the measured phases completely removes laser phase noise. 
\begin{equation}
\label{TDIcomb1}
\begin{gathered}
2([\hat{\phi}^{\text{g}}_{21}(t)-\hat{\phi}^{\text{g}}_{23}(t)]-[\hat{\phi}^{\text{g}}_{21}(t-2\tau_{23})-\hat{\phi}^{\text{g}}_{23}(t-2\tau_{21})]) \;,
\end{gathered}
\end{equation}
where $\hat{\phi}^{\text{g}}_{ij}(t)$ denotes the measured phase shift at time t using light received at satellite $i$ from satellite $j$. Converting the gravitational phase shift to the accelerations of the different satellites (e.g. $\hat{\phi}^{\text{g}}_{21}(t)\rightarrow a_{\text{g},1}(t)-a_{\text{g},2}(t)$) gives the ranging signal as
\begin{equation}
\begin{split}
a_{R(TDI)}(t)=2(&[a_{\text{g},1}(t)-a_{\text{g},2}(t)-(a_{\text{g},2}(t)-a_{\text{g},3}(t))]\\
-&[a_{\text{g},1}(t-2\tau_{23})-a_{\text{g},2}(t-2\tau_{23})\\
-&(a_{\text{g},2}(t-2\tau_{21})-a_{\text{g},3}(t-2\tau_{21}))])\;,
\end{split}
\end{equation}
where $a_{\text{g},i}(t)$ is the range acceleration of the $i$th satellite at time t. Assume that the satellites are all separated by the same distance, so that $\tau_{L}=\tau_{21}=\tau_{23}$, where the subscript L denotes that we are referring to a light time of flight, and $a_{\text{g},3}(t)=a_{\text{g},2}(t-\tau_S)=a_{\text{g},1}(t-2\tau_S)$. Converting to the frequency domain then gives
\begin{equation}
\begin{split}
a_{R(TDI)}(f)=&\text{ }2a(f)[(1-e^{-2\pi \mathrm{i} f \tau_{S}}-(e^{-2\pi \mathrm{i} f \tau_{S}}-e^{-2\pi \mathrm{i} f 2\tau_{S}})) \\ 
&-(e^{-2\pi \mathrm{i} f 2\tau_{L}}-e^{-2\pi \mathrm{i} f (\tau_{S}+2\tau_{L})}-\\
&\text{ }(e^{-2\pi \mathrm{i} f (\tau_{S}+2\tau_{L})}-e^{-2\pi \mathrm{i} f (2\tau_{S}+2\tau_{L})}))]\\
=&\text{ }2a(f)(1-e^{-2\pi \mathrm{i} f \tau_{S}}-(e^{-2\pi \mathrm{i} f \tau_{S}}-e^{-2\pi \mathrm{i} f 2\tau_{S}}))\\
&\times(1-e^{-2\pi \mathrm{i} f 2\tau_{L}}) \\ 
=&\text{ }2a(f)(1-e^{-2\pi \mathrm{i} f \tau_{S}})^2(1-e^{-2\pi \mathrm{i} f 2\tau_{L}}) \;.
\end{split}
\end{equation}
Taking the absolute value of this gives the expression in the main text, Eq.~\eqref{eq:acc_TDIsat}. The acceleration range after TDI is proportional to $\abs{\text{sin}(\pi f/f_c)}$, where $f_c=c/(2L)$. This introduces nulls in the signal at certain frequencies which are not present in the original ranging signal. After rescaling by a factor $1/(2\sqrt{2}\abs{\text{sin}(\pi f/f_c)})$, so that quantum noise is unaffected by TDI, the signal is affected by a $\sqrt{2}\abs{\text{sin}(2\pi f/f_L)}$ term compared to the signal without TDI. The main effect of this is to reduce the signal at low frequencies. Close to the frequency of interest TDI does not affect the signal significantly.

\section{TDI with unequal satellite separations}
\label{apen:unequalsep}
\begin{figure}[t]
\centering
\includegraphics[width=0.47\textwidth]{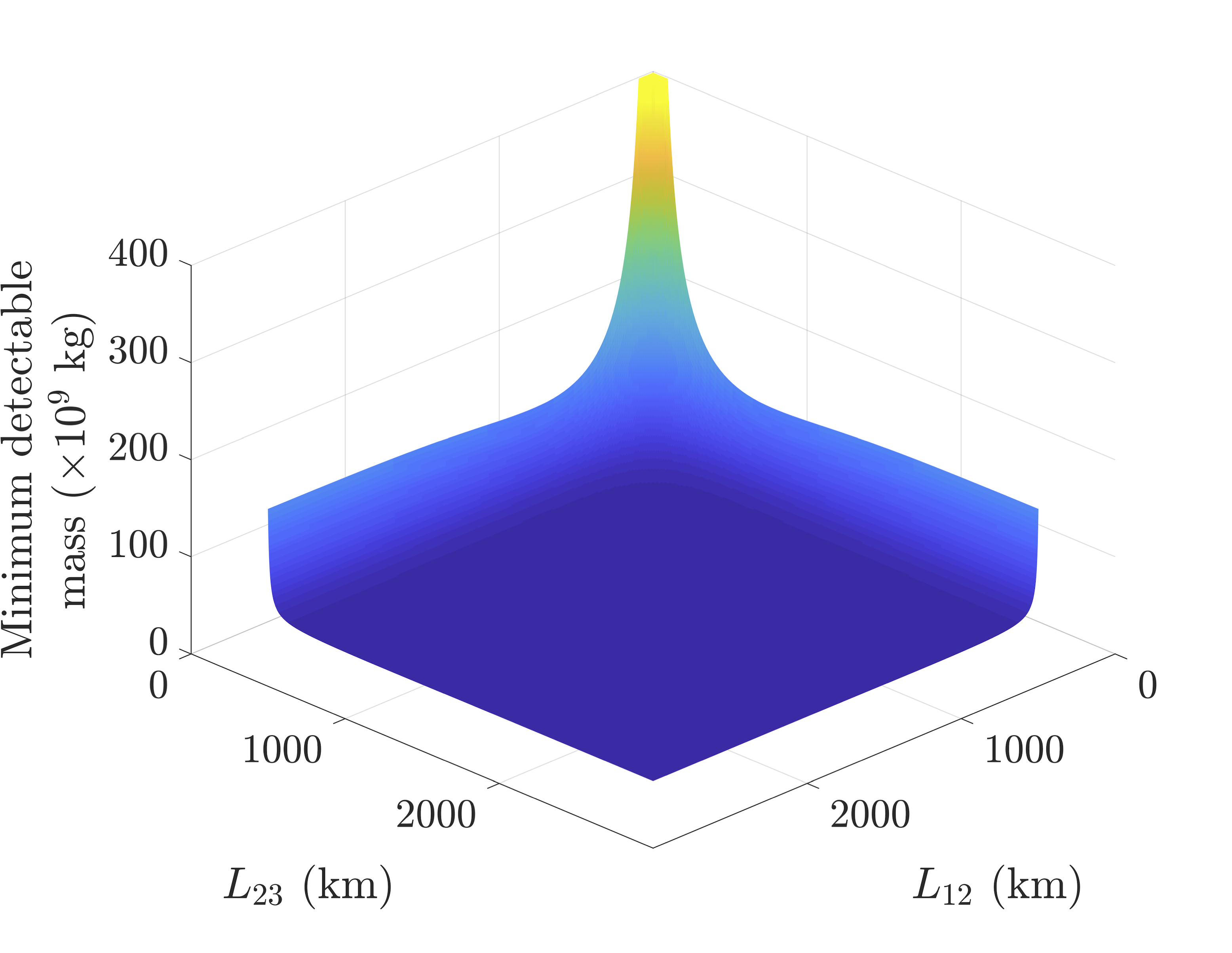}
\caption{\textbf{Minimum detectable mass using TDI with different satellite separations.} The minimum detectable mass is shown as a function of the distance between satellites 1 and 2, $L_{12}$ and between satellites 2 and 3, $L_{23}$, when using time delay interferometry. Figure is shown for satellites at an orbital height of 500 km with an accelerometer noise of $a_0=1\times10^{-12}\text{ m/s}^{2}\sqrt{\text{Hz}}$. Quantum noise is negligible at this level of accelerometer noise. }
\label{fig:diff_sat_length}
\end{figure}

For completeness we now consider the first TDI combination presented, i.e. formation $\alpha^T_3$ with a single laser on the middle satellite, in the situation where the satellite separations are not equal. The acceleration combination which is laser phase noise free is
\begin{eqnarray}
a_{\text{TDI}}&=&a_{g,12}(t)-a_{g,23}(t)-\nonumber\\
&&(a_{g,12}(t-2\tau_{23})-a_{g,23}(t-2\tau_{12}))\nonumber\\
&=&a_{g,1}(t)-a_{g,2}(t)-a_{g,2}(t)+a_{g,3}(t)-\nonumber\\
&& (a_{g,1}(t-2\tau_{23})-a_{g,2}(t-2\tau_{23})\nonumber\\
&&-a_{g,2}(t-2\tau_{12})+a_{g,3}(t-2\tau_{12}))\nonumber\\
&=&a_{g,1}(t)-2a_{g,1}(t-\tau_{S12})+a_{g,1}(t-\tau_{S12}-\tau_{S23})- \nonumber\\
&& (a_{g,1}(t-2\tau_{23})-a_{g,1}(t-\tau_{S12}-2\tau_{23})-\nonumber \\
&&a_{g,1}(t-\tau_{S12}-2\tau_{12})+a_{g,1}(t-\tau_{S12}-\tau_{S23}-2\tau_{12}))\;,\nonumber\\
&&
\end{eqnarray}
where we are using the fact that the accelerations of all the satellites are equal but delayed in time and $\tau_{ij}$ refers to the light time of travel and $\tau_{Sij}$ refers to the satellite time of travel. Taking the Fourier transform we find
\begin{equation}
\begin{split}
a_2(f)=&\text{ }a(f)[(1-e^{-2\pi i f (2\tau_{23})})
\\&+e^{-2\pi i f (\tau_{S12})}(-2-e^{-2\pi i f (2\tau_{23})}+e^{-2\pi i f (2\tau_{12})})\\
&+e^{-2\pi i f (\tau_{S12}+\tau_{S23})}(1-e^{-2\pi i f (2\tau_{12})})]\;.
\end{split}
\end{equation}
where $a(f)$ is the acceleration frequency spectrum of a single satellite, as given in Eq.~\eqref{acc_0} and the subscript 2 refers to the fact that this combination has a laser on-board the middle satellite (satellite 2). When the satellite separations are the same, $\tau_{S12}=\tau_{S23}=\tau_S$ and $\tau_{12}=\tau_{23}=\tau$, then this reduces to $a(f)(1-e^{-2\pi i f(2\tau)})(1-e^{-2\pi i f(\tau_S)})^2$, as expected. Similarly if the laser is at satellite 1 initially, the signal after TDI becomes
\begin{equation}
\begin{split}
a_1(f)=&\text{ }a(f)[e^{-2\pi i f (\tau_{S12}+\tau_{S23})}-e^{-2\pi i f (\tau_{S12})}\\
&-e^{-2\pi i f (2\tau_{12}+2\tau_{23})}+e^{-2\pi i f (\tau_{S12}+2\tau_{23}+2\tau_{12})}\\
&+e^{-2\pi i f (2\tau_{12})}-e^{-2\pi i f (2\tau_{12}+\tau_{S12}+\tau_{S23})}]\;.
\end{split}
\end{equation}
With the laser at satellite 3 the signal becomes
\begin{equation}
\begin{split}
a_3(f)=&\text{ }a(f)[e^{-2\pi i f \tau_{S12}}-1-e^{-2\pi i f (2\tau_{12}+2\tau_{23}+\tau_{S12})}\\
&+e^{-2\pi i f (2\tau_{23})}+e^{-2\pi i f (\tau_{S12}+\tau_{S23}+2\tau_{23}+2\tau_{12})}\\
&-e^{-2\pi i f (2\tau_{23}+\tau_{S12}+\tau_{S23})}]\;.
\end{split}
\end{equation}
The accelerometer noise after TDI, with the laser at the middle satellite, transforms as
\begin{equation}
\begin{split}
AN_{\text{TDI,2}}(f)\rightarrow &AN_{21}(f)(1-e^{-2\pi i f 2\tau_{23}})\\&+AN_{23}(f)(1-e^{-2\pi i f 2(\tau_{12})})\;.
\end{split}
\end{equation}
Similarly if the laser is at satellite 1 or 3 the accelerometer noise transforms as
\begin{equation}
\begin{split}
AN_{\text{TDI,1}}(f)\rightarrow &AN_{21}(f)(1-e^{-2\pi i f 2(\tau_{12}+\tau_{23})})\\&+AN_{13}(f)(1-e^{-2\pi i f 2(\tau_{12})})\;,
\end{split}
\end{equation}
and
\begin{equation}
\begin{split}
AN_{\text{TDI,3}}(f)\rightarrow& AN_{23}(f)(1-e^{-2\pi i f 2(\tau_{12}+\tau_{23})})\\&+AN_{13}(f)(1-e^{-2\pi i f 2(\tau_{23})})\;,
\end{split}
\end{equation}
respectively. For the accelerometer instrument noise, the various $AN_{i,j}$ terms above will be statistically similar. However, for the quantum noise as the two arms are different lengths the quantum noise levels in the two measurements will be different, hence the corresponding $QN_{i,j}$ terms will not be statistically similar. This allows the minimum detectable mass to be examined as a function of both satellite separations, as shown in Fig.~\ref{fig:diff_sat_length}. However, this does not reveal any interesting new optimal regimes for satellite geodesy. As expected, we see that the minimum detectable mass is symmetric in both arm lengths.

\section{Experimental aspect of TDI}
\label{LPNexp}
In principle TDI allows for the perfect removal of laser phase noise. This relies on time-delaying the data series by a time corresponding exactly to the time of flight of light travelling between the two satellites. However, in reality the satellite arm length will not be known perfectly, rather it will be known to within some uncertainty, $\hat{\tau}_{ij}=\tau_{ij}+\delta_j$, where $\hat{\tau}_{ij}$ is the estimated time of flight and $\delta_j$ is a random variable which represents the error in this estimate~\cite{tinto2014time}. Substituting the estimated time of flight into the TDI combinations shows that without perfect arm length knowledge there will be some left-over laser phase noise. For formation $\alpha^T_3$ with a single laser on the middle satellite, the left-over laser phase noise is 
\begin{equation}
\begin{split}
\text{LPN}_{\text{left-over}}=&\text{ }C_2(t-2\tau_{21})-C_2(t-2\tau_{21}-2\delta_1)\\
&-C_2(t-2\tau_{23})+C_2(t-2\tau_{23}-2\delta_3) \\
& +C_2(t-2\tau_{21}-2\tau_{23}-2\delta_{3})\\
&-C_2(t-2\tau_{21}-2\tau_{23}-2\delta_1)\;.
\end{split}
\end{equation}
In order to calculate the remaining laser phase noise after TDI we expand about $\delta_i=0$ to first order and assume the uncertainty in both arm lengths is the same, $\delta_1=\delta_3=\delta$. In the time domain this gives
\begin{equation}
\text{LPN}_{\text{left-over}}=\text{ }2\delta(\dot{C}_2(t-2\tau_{21})-\dot{C}_2(t-2\tau_{23}))\;.
\end{equation}
Converting to the frequency domain we obtain the new laser phase noise spectrum. 
\begin{equation}
\sqrt{S_{\text{LPN}_{\text{left-over}}}(f)} \approx8\pi\delta f\sqrt{S_{\text{LPN}}} \;.
\end{equation}

This is the approach we use for the simulations in the main text. However, as an alternative approach which may be more intuitive, we can note that the model being used for laser phase noise is linear in satellite separation. Assuming this holds for small distances, we can simply look at the total \lq distance\rq which remains in the expression for left-over laser phase noise, i.e. $4\delta_1+4\delta_3$. A pessimistic estimate for the left-over laser phase noise amplitude spectrum is then
\begin{equation}
\sqrt{S_{\text{LPN}_{\text{left-over}}}(f)}=\frac{(2\pi f)^2x_T8 \delta}{\sqrt{f}}\;,
\end{equation}
where we assume that $\delta_1\approx\delta_3$ and $\delta=c\delta_1$ is a distance as opposed to a time. The requirement to be quantum noise limited when using this TDI combination is that the power spectrum of the left-over laser phase noise is smaller than that of the quantum noise spectrum, $\abs{\sqrt{S_{\text{LPN}_{\text{left-over}}}(f)}}\leq\sqrt{S_{\text{QN,TDI}}(f)}$. There is a similar requirement on the accelerometer noise, in order for the left-over laser phase noise to be unimportant. This is more relevant to the present mission as accelerometer noise is larger than quantum noise. This places a requirement on how accurately the arm lengths must be known for TDI to be beneficial. With GPS positioning accuracy on the order of 5 mm~\cite{kroes2005precise,wu2006real}, this requirement is easily surpassed. Millimetre level positioning is sufficiently accurate for TDI not to be a limiting factor even for $a_0=1\times10^{-15}\text{ m/s}^2\sqrt{\text{Hz}}$. Thus TDI offers a practical and attainable method of improving the sensitivity of geodesy missions.

\section{Noise sources after TDI}
\label{tdi_noise_sources_apen}
When taking the above TDI combination several noise terms are combined in the final expression. The different noise terms combine in different ways depending on whether they are correlated or not. We consider quantum noise first. Adding two uncorrelated quantum noise terms, is adding two series of time data with a certain variance, $\sigma_1^2$ and $\sigma_2^2$. The total variance is $\sigma_{\text{T}}^2=\sigma_1^2+\sigma_2^2$. This has power spectral density proportional to $\sigma_{\text{T}}^2$ and so amplitude spectral density proportional to $\sigma_{\text{T}}=\sqrt{\sigma_1^2+\sigma_2^2}$. Assuming the two quantum noise spectra being combined are statistically similar (same variance) we see that $\sigma_{\text{T}}=\sqrt{2}\sigma_1$, i.e. amplitude spectra get scaled by a factor $\sqrt{2}$. Adding $N$ quantum noise terms together will scale the total quantum noise amplitude spectrum by a factor of $\sqrt{N}$, compared to each individual spectrum (assuming all the spectra are statistically similar). 

This is true for quantum noise which has a white noise spectrum. However, the above argument is easily extended to any shape of frequency spectrum. Any noise spectrum, $A(f)$, has a shape which depends on the frequency. This function determines the magnitude of the noise at that frequency. In order to obtain a noisy frequency spectrum the magnitude of the spectrum at each frequency can be multiplied by a normally distributed random number with mean 0 and variance 1, $\mathcal{N}(0,1)$, to change its magnitude and multiplied by a random complex number, $e^{\mathrm{i}\theta}$, where $\theta$ is distributed uniformly in the region $0$ to $2\pi$, to change its phase. We then consider adding $j$ of these similar noise spectra together. At a given frequency, $f$, we can model the total frequency spectrum as
\begin{equation}
\abs{A(f)}\sum_je^{\mathrm{i}\theta_j}\mathcal{N}_j(0,1)\;.
\end{equation}
The absolute value of this spectrum then indicates how much noise we can expect at a given frequency. As the spectral shape $A(f)$ is outside the sum, at any given frequency the expected value of this is simply $\sqrt{j}$ times larger than a single noise spectra. This is true at all frequencies and for any spectral shape. 

We can now use this to calculate how the quantum noise or accelerometer noise spectrum is affected by TDI. From Eq.~\eqref{TDIcomb} in the main text, the quantum noise in the time domain after taking the TDI combination is 
\begin{equation}
\begin{split}
QN_{\text{TDI}}(t)=&\text{ }(QN_{21}(t)-QN_{21}(t-2\tau_{23}))\\
&-(QN_{23}(t)-QN_{23}(t-2\tau_{21}))\;,
\end{split}
\end{equation} 
where $QN_{ij}(t)$ ($AN_{ij}(t)$) represent the quantum noise (accelerometer noise) for a measurement using light received at satellite $i$ from satellite $j$. Converting to the frequency domain gives
\begin{equation}
\begin{split}
QN_{\text{TDI}}(f)=&\text{ }QN_{21}(f)(1-e^{-2\pi \mathrm{i} f (2\tau_{23})})\\
&+QN_{23}(f)(1-e^{-2\pi \mathrm{i} f (2\tau_{21})})\;,
\end{split}
\end{equation}
which, assuming that the two quantum noise spectra are statistically equivalent and that the satellites separations are initially equal, gives
\begin{equation}
\abs{QN_{\text{TDI}}(f)}=2\sqrt{2}QN(f)\abs{\text{sin}(\frac{2\pi f}{f_c})}\;.
\end{equation}
The accelerometer noise is transformed in the same way. The remaining laser phase noise is in theory perfectly cancelled. However, with experimental imperfections, this is not the case, as discussed in appendix~\ref{LPNexp}.

\section{Minimum detectable mass}
\label{MDM_apen}
\begin{figure*}[t]
\includegraphics[width=0.8\textwidth]{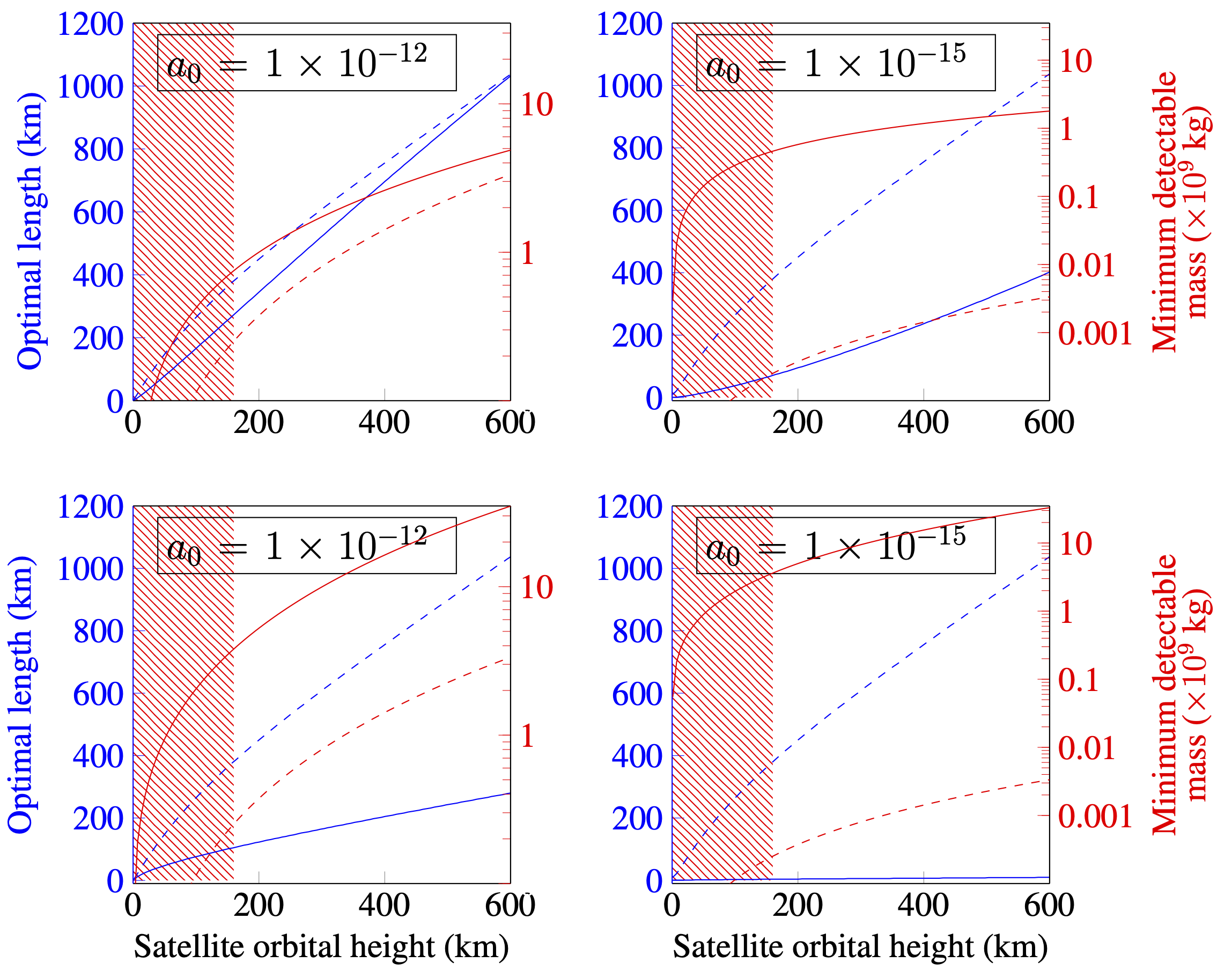}
\caption{\textbf{Minimum detectable mass and corresponding optimal satellite separation for different satellite orbital heights.} The data is shown for two different accelerometer instrument noise levels, $a_0=1\times10^{-12}\text{ m/s}^{2}\sqrt{\text{Hz}}$ and $a_0=1\times10^{-15}\text{ m/s}^{2}\sqrt{\text{Hz}}$ both with (dashed lines) and without (solid lines) TDI. The hashed red region represents satellite orbital heights below low Earth orbit (160 km) and so this region is not feasible. The top and bottom rows show the same data but using two different initial laser phase noise spectra, corresponding to the actual laser phase noise performance ($x_T=1\times10^{-15}$, top row) and the laser phase noise requirement ($x_c=8\times10^{-9}$ m/$\sqrt{\text{Hz}}$, bottom row) respectively.}
\label{fig:MDM_sat_height}
\end{figure*}
The minimum detectable mass quantifies the useful region of the frequency spectrum of the gravitational signal and all the noise sources. When using a matched filter, which is optimal if the shape of the signal is known, the SNR of the gravitational ranging system for detecting a point mass M is~\cite{flanagan1998measuring}
\begin{equation}
\rho=4\int_{0}^{\infty}\frac{\abs{a_{\text{R}}(f)}^2}{S_{\text{T}}(f)}df\;,
\label{SNR}
\end{equation}
where $a_{\text{R}}(f)$ is the Fourier transform of the time domain gravitational signal and $S_{\text{T}}(f)$ is the combined power spectrum of all the different noise sources in the system. Hence the SNR per unit mass is
\begin{equation}
\rho'=4\int_{0}^{\infty}\frac{\abs{a_{\text{R}}(f)/M}^2}{S_{\text{T}}(f)}df\;.
\label{SNRpum}
\end{equation}
In the main text the minimum detectable mass was defined in Eq.~\eqref{minmass} as
\begin{equation}
M_{\text{min}}=\frac{3}{\sqrt{4\int_{0}^{\infty}\frac{\abs{a_{\text{R}}(f)/M}^2}{S_{\text{T}}(f)}df}}\;.
\end{equation}
It is now clear that the minimum detectable mass is simply the smallest mass which gives a SNR of at least 3. Intuitively this concept represents the smallest possible mass which the satellite system can detect. This concept allows the optimal satellite separation (that which minimises the minimum detectable mass) to be determined for a given orbital height. In the main text this was presented for satellites at an orbital height of 500 km. This orbital height was chosen as it is the orbital height of the current GRACE mission. We now present the optimal satellite separation, and the corresponding minimum detectable mass, as a function of satellite orbital height in Fig.~\ref{fig:MDM_sat_height}. The advantage of improving the accelerometer instrument noise by 3 orders of magnitude is very marginal without TDI. This is because without TDI, laser phase noise remains the dominant noise source and so improving the accelerometer does not help.

\section{Alternative accelerometer noise free combinations}
\label{AltAccComb}
In the main text we briefly mentioned how accelerometer noise free combinations can be obtained from TDI combinations using three satellites of different masses ($\text{M}_1$, $\text{M}_2$ and $\text{M}_3$), formation $\alpha^T_{3,DM}$ in the main text. This formation works based on the assumption that the non-gravitational accelerations experienced by the satellites consists of a stationary, $a^{\text{ng}}_{\text{s}}$, and a non-stationary, $a^{\text{ng}}_{\text{ns}}$, component. All satellites experience the same stationary component and the closer the satellites are to each other the more similar the non-stationary components they experience will be. We can formalise this as $a^{\text{ng}}_{2}(t)=a^{\text{ng}}_{1}(t-\tau_s)+\delta_{a2}$(t), where $a^{\text{ng}}_{i}$ is the non-gravitational acceleration experienced at satellite i, $\delta_{a i}$ is the difference between the non-gravitational acceleration of the first satellite and the non-gravitational acceleration of satellite $i$ at the same position and $\tau_s$ is the time it takes the second satellite to reach the position of the first one. Calling $L_{i}$, the single laser TDI combination with a laser at the ith satellite (i.e. $L_{2}(t)$ is given in Eq.~\eqref{TDIcomb} in the main text), the following combinations of TDI combinations cancel the stationary component of the accelerometer noise
\begin{equation}
\begin{split}
\text{Com}_1=&\text{ }(L_{1}(t)-L_{1}(t-2\tau_{23})+ \frac{M_1}{M_3}L_{1}(t-2\tau_{S})\\
&-\frac{M_1}{M_3} L_{1}(t-2\tau_{21}-2\tau_{S})-2\frac{M_1}{M_2} L_{1}(t-\tau_S) \\
&+\frac{M_1}{M_2} L_{1}(t-2\tau_{21}-\tau_S) +\frac{M_1}{M_2} L_{1}(t-2\tau_{23}-\tau_S)) \\
& - [L_{2}(t-2\tau_{21})-L_{2}(t-2\tau_{21}-2\tau_{23}) \\
&+\frac{M_1}{M_3} L_{2}(t-2\tau_S)-\frac{M_1}{M_3} L_{2}(t-2\tau_{21}-2\tau_S) \\
&-\frac{M_1}{M_2} L_{2}(t-\tau_S)+\frac{M_1}{M_2} L_{2}(t-2\tau_{21}-2\tau_{23}-\tau_S)]\;.
\end{split}
\end{equation}

\begin{equation}
\begin{split}
\text{Com}_2=&\text{ }(-L_{1}(t)+L_{1}(t-2\tau_{23})-\frac{M_1}{M_3} L_{1}(t-2\tau_{S}-2\tau_{23})\\
&+\frac{M_1}{M_3} L_{1}(t-2\tau_{21}-2\tau_{S}-2\tau_{23})+\frac{M_1}{M_2} L_{1}(t-\tau_S)\\
&-\frac{M_1}{M_2} L_{1}(t-2\tau_{21}-2\tau_{23}-\tau_S))+\frac{M_1}{M_3} L_{3}(t-2\tau_S) \\
&- [L_{3}(t-2\tau_{21})-L_{3}(t-2\tau_{21}-2\tau_{23})\\
&-\frac{M_1}{M_3} L_{3}(t-2\tau_{21}-2\tau_S)-\frac{M_1}{M_2} L_{3}(t-\tau_S)\\
&+\frac{M_1}{M_2} L_{3}(t-2\tau_21-2\tau_23-\tau_S)]\;,
\end{split}
\end{equation}

\begin{equation}
\begin{split}
\text{Com}_3=&\text{ }(L_{3}(t)-L_{3}(t-2\tau_{23})+\frac{M_1}{M_3} L_{3}(t-2\tau_{S})\\
&-\frac{M_1}{M_3} L_{3}(t-2\tau_{21}-2\tau_{S})-2\frac{M_1}{M_2} L_{3}(t-\tau_S)\\
&+\frac{M_1}{M_2} L_{3}(t-2\tau_{21}-\tau_S)+\frac{M_1}{M_2} L_{3}(t-2\tau_{23}-\tau_S)) \\ 
&-[-L_{2}(t)+L_{2}(t-2\tau_{23})-\frac{M_1}{M_3} L_{2}(t-2\tau_{S}-2\tau_{23})\\
&+\frac{M_1}{M_3} L_{2}(t-2\tau_{21}-2\tau_{S}-2\tau_{23})+\frac{M_1}{M_2} L_{2}(t-\tau_S)\\
&-\frac{M_1}{M_2} L_{2}(t-2\tau_{21}-2\tau_{23}-\tau_S)]\;.
\end{split}
\end{equation}
As the laser phase noise is already removed in the TDI combinations used here, $L_1$, $L_2$ and $L_3$, these terms only have quantum noise and non-stationary non-gravitational accelerations remaining. However, there are many issues with these combinations; engineering difficulties, greatly reduced signal and the remaining noise sources. As such these schemes appear to have limited practical value.

\section{Remaining noises with six satellite formation flying}
\label{SixSatApen}
\begin{figure*}[t]
\includegraphics[width=\textwidth]{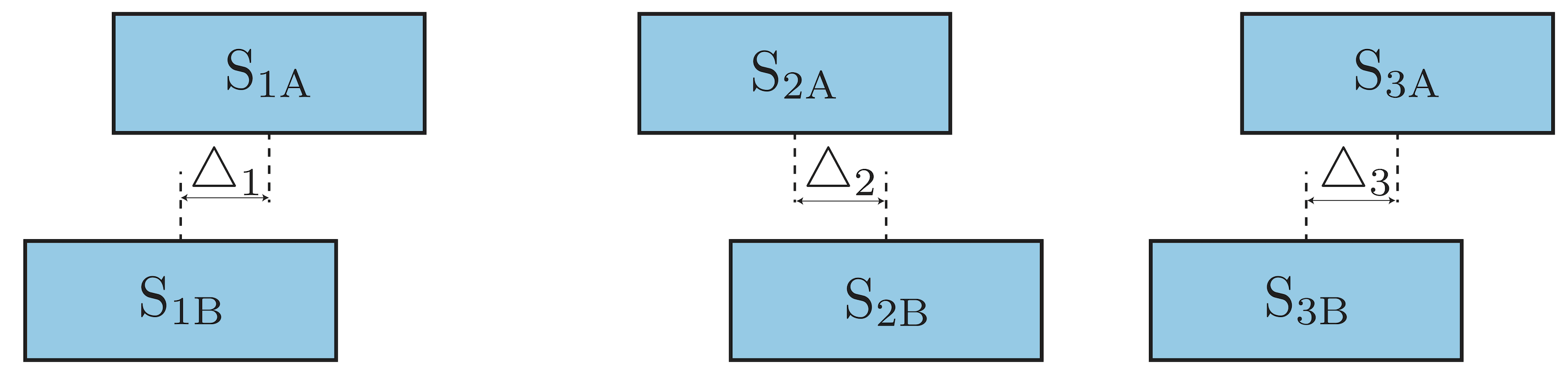}
\caption{\textbf{Satellite drift for the six satellite formation presented in the main text, formation $\alpha^T_{6,DM}$.} The three satellite-pairs drift by distances $\triangle_{1}$, $\triangle_{2}$ and $\triangle_{3}$, which are all of a similar order of magnitude. The smaller the satellite-pair drift can be made the better the performance of this scheme. The inter-satellite separation used for the ranging measurement is on the order of hundreds of kilometres.}
\label{fig:sixsatdrift}
\end{figure*}
The differential mass 6 satellite formation presented in the main text, formation $\alpha^T_{6,DM}$, exactly cancels the non-gravitational accelerations and the laser phase noise in an ideal world. However, in reality each satellite pair will drift apart. If the three satellite pairs drift apart by distances $\triangle_{1}$, $\triangle_{2}$ and $\triangle_{3}$, as shown in Fig.~\ref{fig:sixsatdrift}, the remaining laser phase noise is
\begin{equation}
\begin{split}
\text{LPN}_{\text{left-over}}=&\text{ }2(C_2(t-2\tau_{21}+2\triangle_{t1})-C_2(t-2\tau_{21}+2\triangle_{t2})\\
&-C_2(t-2\tau_{23}+2\triangle_{t3})+C_2(t-2\tau_{23}+2\triangle_{t2})\\
&-C_2(t-2\tau_{21}-2\tau_{23}+2\triangle_{t1})\\
&+C_2(t-2\tau_{21}-2\tau_{23}+2\triangle_{t3}))\;,
\end{split}
\end{equation}
where $\triangle_{ti}=\triangle_i/v_0$, is the time corresponding to each satellite drift. Following the same logic as used in appendix~\ref{LPNexp}, this leads to the following expression for the remaining laser phase noise, assuming all the satellites drift a similar distance, $\triangle_{i}$=$\triangle$,
\begin{equation}
\sqrt{S_{\text{LPN}}(f)}=\frac{(2\pi f)^2x_T12\triangle}{\sqrt{f}}\;.
\end{equation}
This however assumes that even if the satellites drift apart we take the same naive TDI combination. Alternatively we can take a different TDI combination which is able to almost completely remove laser phase noise. We form the two quantities $\phi_{T21}(t)$ and $\phi_{T23}(t)$ as described in the main text ($2\phi_{T2i}(t)=4\hat{\phi}_{B2i}(t)-2\hat{\phi}_{A2i}(t)$). The laser phase noise in these quantities is
\begin{equation}
\begin{split}
C_{T21}(t)=&\text{ }2C(t-2\tau_{21}+\triangle_{t1})-C(t-2\tau_{21})\\
&-2C(t+\triangle_{t2})+C(t)\;,
\end{split}
\end{equation}
and
\begin{equation}
\begin{split}
C_{T23}(t)=&\text{ }2C(t-2\tau_{23}+\triangle_{t3})-C(t-2\tau_{23})\\
&-2C(t+\triangle_{t2})+C(t)\;.
\end{split}
\end{equation}
The following combination is laser phase noise free
\begin{equation}
\begin{split}
\phi_{\text{6,T}}(t)=&\text{ }\phi_{T21}(t)-\phi_{T23}(t)\\
&-[\phi_{T21}(t-2\tau_{23})-\phi_{T23}(t-2\tau_{21})]\\
&+2[\phi_{T21}(t-2\tau_{23}+\triangle_{t3})-\phi_{23}(t-2\tau_{21}+\triangle_{t1})]\\
&-2[\phi_{T21}(t+\triangle_{t2})-\phi_{T23}(t+\triangle_{t2})]\;,
\end{split}
\end{equation}
where we use the term $\phi_{\text{6,T}}$ to reflect that this corresponds to formation $\alpha^T_{6,DM}$. As in appendix~\ref{LPNexp}, including errors in our knowledge of the satellite separations allows an approximate model for the laser phase noises remaining after taking this combination to be calculated. Doing this indicates that the remaining laser phase noise in this six satellite configuration will be no more than an order of magnitude larger than the laser phase noise left over in our three satellite configuration. Remarkably, even though there are many more terms in this TDI combination than the more simple TDI configuration initially chosen, the frequency domain signals of the two combinations are very similar. In the frequency domain this signal has the following form
\begin{equation}
\begin{split}
a_{\text{6,TDI}}(f)=&\text{ }a_R(f)[1-e^{-2\pi i f \tau_S}\\
&-[e^{-2\pi i f2\tau_L}-e^{-2\pi i f(2\tau_L+\tau_S)}]
\\&+2[e^{-2\pi i f(2\tau_L-\triangle_{t3})}-e^{-2\pi i f(2\tau_L-\triangle_{t1}+\tau_S)}]\\
&-2[e^{-2\pi i f(-\triangle_{t2})}-e^{-2\pi i f(\tau_S-\triangle_{t2})}]]\;,
\end{split}
\end{equation}
where $a_R(f)$ is the range acceleration signal given by Eq.~\eqref{eq:arsig} and as before, $\tau_L$ and $\tau_S$ are the single trip time of flight for light and the satellites respectively. When $\triangle_1=\triangle_2=\triangle_3=0$, this reduces to the normal TDI signal. When this TDI combination is taken laser phase noise is almost completely removed. However, the other noise sources remain in the measurement. When the individual measurements are combined, $\phi_{T i,j}=2\phi_{B i, j}-\phi_{A i,j}$, the quantum noise in the $\phi_{B i, j}$ term is doubled. This combined with the noises in $\phi_{A i,j}$ (which we assume to be statistically equivalent) means the quantum noise in the $\phi_{T i,j}$ term is a factor of $\sqrt{5}$ larger than in the $\phi_{A(B) i, j}$ terms. In the final TDI expression there are then two independent quantum noise terms (one each from $\phi_{T21}(t)$ and $\phi_{T23}(t)$). For this TDI combination the quantum noise is transformed as    
\begin{equation}
\begin{split}
QN(f)\rightarrow &\text{ }QN_{21}'(f)[1-e^{-2\pi i f2\tau_L}\\
&+2e^{-2\pi i f (2\tau_{L}-\triangle_{t3})}-2e^{2\pi i f \triangle_{t2}}] \\
 &+QN_{23}'(f)[-1+e^{-2\pi i f2\tau_L}\\
 &-2e^{-2\pi i f (2\tau_{L}-\triangle_{t1})}+2e^{2\pi i f \triangle_{t2}}]\;,
\end{split}
\end{equation}
where we have assumed the satellite separations are initially equal and $QN_{ij}'(f)$ represents the quantum noise spectrum in the $\phi_{T i,j}$ terms, i.e. the normal quantum noise scaled by $\sqrt{5}$. There is some accelerometer noise left in each $\phi_{T i,j}$ term, owing to imperfect satellite flying, which transforms in a similar way. The combination of all of these left-over noises gives the total noise spectrum for this formation.

\section{Estimating phase when quantum noise limited}
\label{apen:phaseestSNL}

Naively one might imagine that the problem of computing the ultimate precision in satellite geodesy is a typical phase estimation problem~\cite{dorner2009optimal}. Upon delving deeper into the problem it becomes apparent that this is not true. Primarily this is due to the competing noise sources, laser phase noise and accelerometer noise which are both significantly larger than quantum noise. This means that techniques which typically aid phase estimation through the reduction of quantum noise won't help satellite geodesy in its current form. However, at some point in the future such missions may be quantum noise limited. We now numerically investigate this regime with a full 3D model to support the results from our 1D model presented in the main text.
 
 A key difference between a quintessential phase estimation problem and satellite geodesy is that in phase estimation we typically wish to estimate a single number, which is easily extracted from the measurement results. However, in satellite geodesy the quantity of interest is much more complex. The Earth's gravitational potential is normally written as an expansion of the spherical harmonics. 
\begin{equation}
\begin{split}
\label{eq:grav_pot}
V(r,\theta,\phi,t)=&\text{ }\frac{\mu}{r}+\frac{\mu}{r}\sum_{l=2}^{\text{N}_{\text{max}}}\left(\frac{a_e}{r}\right)^l\bar{P}_{lm}(\text{sin}(\theta))\\
&\times[\bar{C}_{lm}(t)\text{cos}(m\phi)+\bar{S}_{lm}(t)\text{sin}(m\phi)]\;,
\end{split}
\end{equation}
where $\theta$ and $\phi$ are latitude and longitude respectively, $r$ is the distance from the satellite to the Earth's centre of mass, $\mu$ is the gravitational constant of the Earth, $a_e$ is the mean equatorial radius of the Earth, $\bar{P}_{lm}(\text{sin}(\theta))$ are the fully normalised associated Legendre polynomials of degree $l$ and order $m$, and $\bar{C}_{lm}(t)$ and $\bar{S}_{lm}(t)$ are the fully normalised spherical harmonic coefficients of the Earth's gravitational potential. The time dependent spherical harmonic coefficients is what the GRACE-FO mission estimates. 

\begin{figure}[t]
\includegraphics[width=0.47\textwidth]{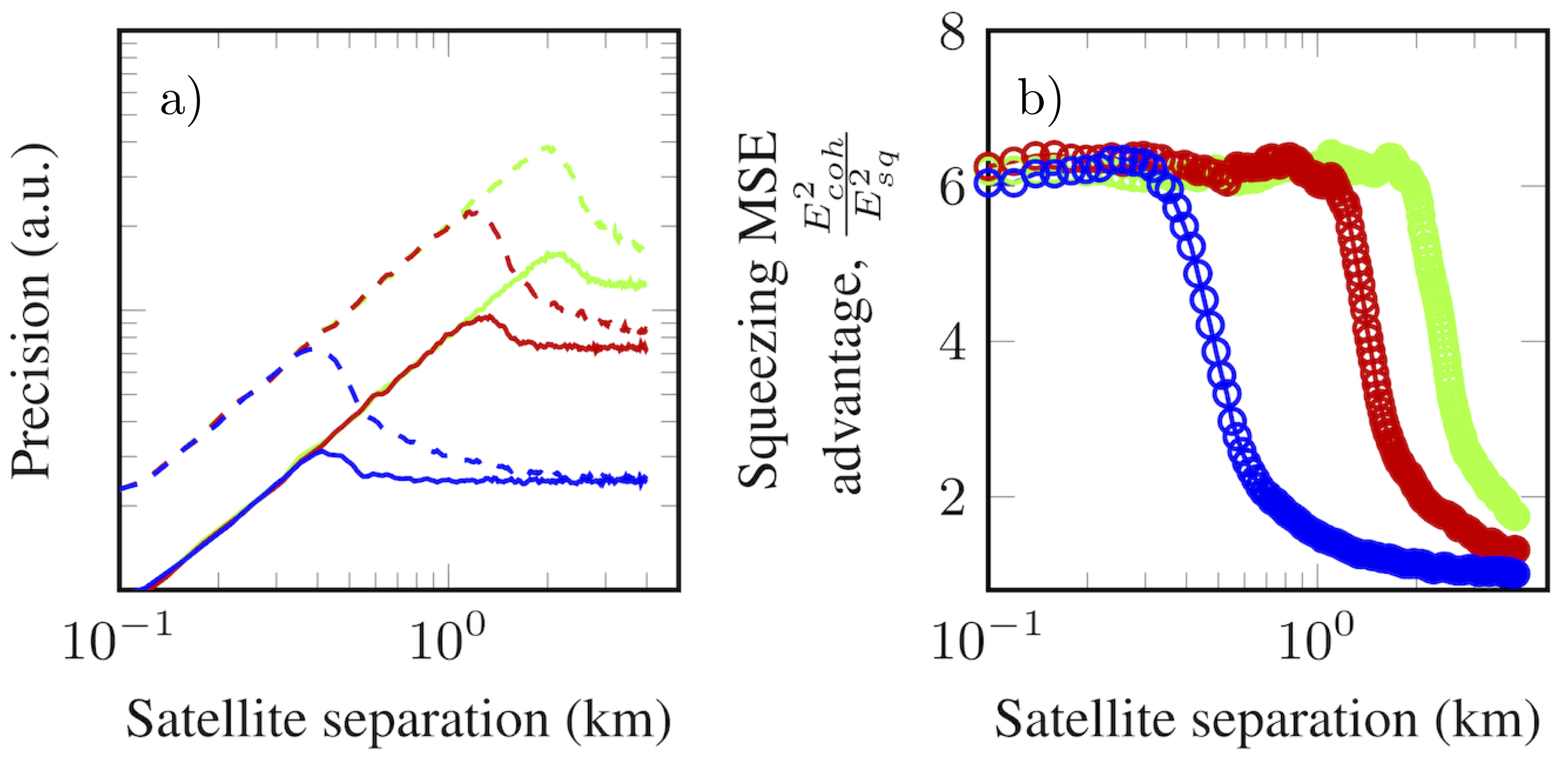}
\caption{\textbf{Numerically attained precision for estimating the $C_{55}$ coefficient of the Earth's gravitational field as a function of satellite separation for different receiving aperture radii}. \textit{Left} shows the achievable precision for several different receiving aperture radii both with and without squeezed light. \textit{Right} shows how the advantage of using squeezed light for satellite geodesy is restricted to a small region, which can be extended by increasing the receiving aperture radius. These simulations assume quantum noise is the dominant noise source. Green, red and blue lines correspond to 25 cm, 15 cm and 5 cm receiving aperture radii respectively. Dashed lines in the figure on the left correspond to using squeezed light.}
\label{fig_3Dapersq}
\end{figure}

In order to numerically verify our analytic results from the 1D model we simplify the problem by estimating only a single coefficient of the Earth's gravitational field, i.e. one $\bar{C}_{lm}(t)$ or $\bar{S}_{lm}(t)$term. We consider a pair of satellites flying in a potential governed by Eq.~\eqref{eq:grav_pot}. It is assumed that we have a prior model of the Earth's gravitational field, i.e. a set of known coefficients, $\bar{C}_{lm}$ and $\bar{S}_{lm}$. The motion of a pair of satellites is simulated in this known potential. From this model we then vary one coefficient by a small amount, approximately $1\%$ and numerically calculate the motion of the satellite pair in the new unknown potential. Quantum noise is added to the true motion of the satellites to give the measured range and based on this we perform a least squares fitting to update our model with a new estimate of $\bar{C}_{lm}$. A major simplification which we make is that we know which coefficient has changed. We define the error as $E=\abs{\bar{C}_{lm}-\hat{\bar{C}}_{lm}}/\bar{C}_{lm}$ and precision as the inverse of the error, where $\hat{\bar{C}}_{lm}$ is the estimate of the updated coefficient. As we are assuming quantum noise is the limit, the receiving aperture size, $a$, plays a key role in determining the achievable precision and optimal satellite separation as shown in Fig.~\ref{fig_3Dapersq}. As predicted in the main text using our 1D model, in the quantum noise limited regime the optimal satellite separation occurs at the point where diffraction loss first becomes significant. In this regime squeezed light can offer a major advantage. The advantage from using squeezing, shown in terms of reduction in mean squared error (MSE), is assuming that  mHz squeezing is available. Although a high squeezing level at mHz is currently unattainable on Earth, this may be easier to achieve in space due to the absence of seismic noise.

\begin{figure}[t]
\centering
\includegraphics[width=0.35\textwidth]{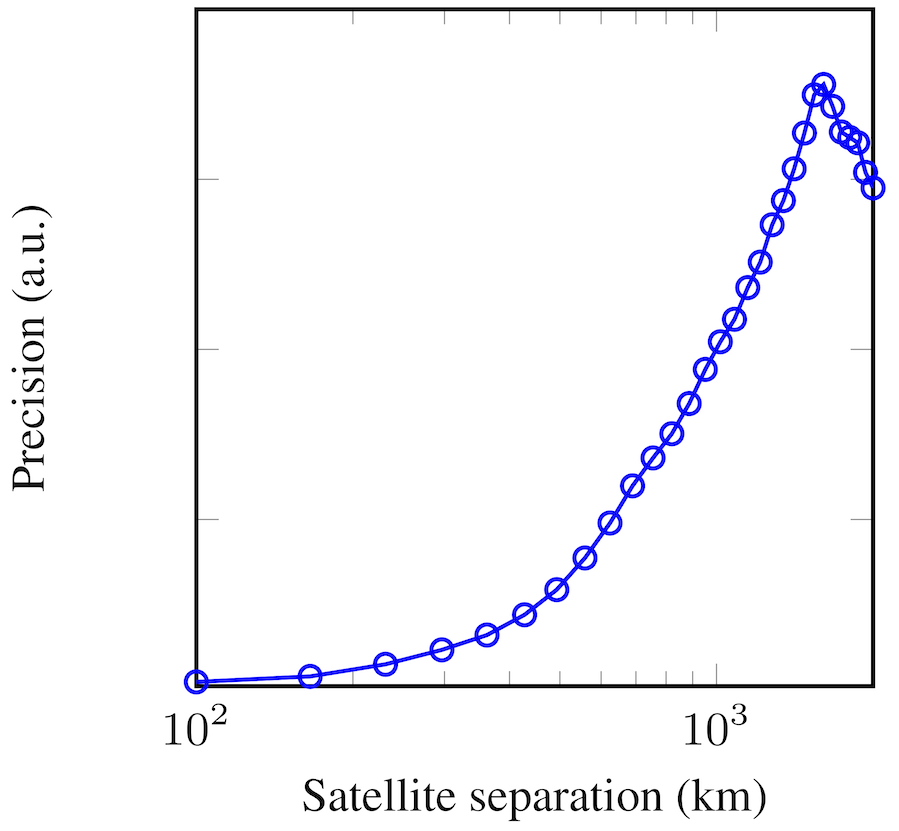}
\caption{\textbf{Attainable precision as a function of satellite separation when using time delay interferometry.} If the quantum noise limit is reached through TDI the optimal satellite separation is considerably larger than if this limit is reached through improvements in laser stability. The y-axis shows the relative precision in estimating the $C_{55}$ coefficient. Reaching the quantum noise limit in this way implies that techniques from quantum optics will not aid gravitational field recovery. This figure assumes that the accelerometer noise is negligible, so that after TDI has been implemented, quantum noise is the dominant noise source.
\label{Fig:TDI_SNlim}}
\end{figure}
We next show that, as predicted by the 1D model in the main text, the way in which the quantum noise limit is reached plays a key role in determining the optimal satellite separation. Prior to now, we have assumed that the quantum noise limit is reached through instrumentation improvement such as an increase in laser stability and a reduction in accelerometer noise. However, if the quantum noise limit is reached using TDI then the optimal satellite separation is very different. Fig.~\ref{Fig:TDI_SNlim} shows the relative precision for estimating the $C_{55}$ coefficient of the gravitational field as a function of satellite separation when TDI is employed. Using TDI the optimal separation is much larger than when TDI is not used and is considerably larger than the region in which squeezed light is useful. The reason for this is that TDI will strongly attenuate any gravitational signal when the satellite separations are small. Hence the benefits of reducing the quantum noise are outweighed by the reduced signal, and the optimal strategy is a large satellite separation. TDI represents the most realistic pathway to reaching the quantum noise limit, further strengthening the argument that squeezed light may not benefit satellite geodesy for the foreseeable future.

\section{Requirements to reach the quantum noise limited regime}
\label{apen:SNLreq}
Finally, we discuss the technological improvements required before the techniques mentioned in the preceding section can be useful, i.e. what are the requirements on laser phase noise and accelerometer noise so that reducing the quantum noise is beneficial. In order to investigate this we consider laser phase noise and accelerometer noise separately, varying $x_T$ and $a_0$, which characterise the two noise sources respectively. 

For a given satellite orbital height and separation we compare the following two terms
\begin{equation}
\label{eq:req1}
\int_0^{\infty}\sqrt{S_{\text{LPN}}}\abs{a_R(f)}df\qquad\text{  or  }\qquad\int_0^{\infty}\sqrt{S_{\text{AN}}}\abs{a_R(f)}df\;,
\end{equation}
and
\begin{equation}
\label{eq:req2}
\int_0^{\infty}\sqrt{S_{\text{QN}}}\abs{a_R(f)}df\;.
\end{equation}

\begin{figure}[t]
\includegraphics[width=0.49\textwidth]{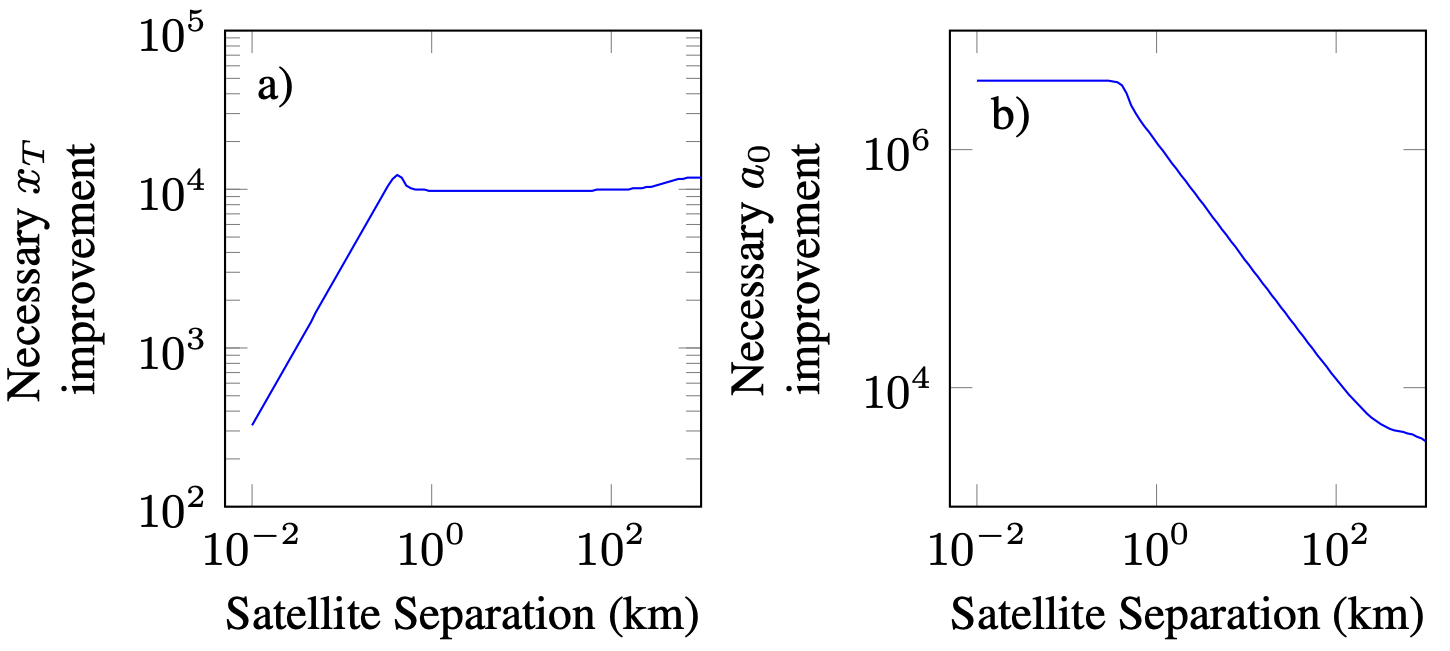}
\caption{\textbf{Laser phase noise and accelerometer noise improvements required to reach the quantum noise limited regime.} Assuming no squeezing, an aperture radius of 5 cm, transmitted laser power of 25 mW, $f_k=5\times10^{-3}\text{ Hz}$ for the accelerometer and a satellite orbital height of 500 km, we show the improvement in laser phase noise and accelerometer noise required to reach the quantum noise limited regime. We assume laser thermal noise $x_T=1\times10^{-15}$ and accelerometer instrument noise $a_0=1\times10^{-12}\text{ m/s}^2\sqrt{\text{Hz}}$ for the comparison.}
\label{figreqSQ}
\end{figure}

The terms $\sqrt{S_{\text{LPN}}}$ and $\sqrt{S_{\text{AN}}}$ depend on the laser thermal noise $x_T$, and accelerometer instrument $a_0$, respectively. For the current GRACE-FO mission the terms in Eq.~\eqref{eq:req1} are considerably larger than that in Eq.~\eqref{eq:req2}. However, through improvements in the laser phase noise and accelerometer noise it is possible to reach the quantum noise limit. The value of $x_T$ or $a_0$ for which the terms in Eq.~\eqref{eq:req1} become equal to the term in Eq.~\eqref{eq:req2} is taken as the region when we are quantum noise limited, assuming a 5 cm receiving aperture radius, initial power, $P_0=25$ mW and squeezing parameter, $r=0.8$. The above expressions look at the noise in the frequency range of interest. From Fig.~\ref{figreqSQ}, we can see that up to 7 orders of magnitude improvement are needed in accelerometer instrument noise before quantum noise becomes a consideration (compared with the projected accelerometer noise for the next GRACE mission, $a_0=1\times10^{-12}\text{m/s}^2\sqrt{\text{Hz}}$). The requirements on laser phase noise are less stringent. The requisite improvements for laser phase noise at small satellite separations are feasible, being approximately 3 orders of magnitude. This hints that there may be an alternative regime for satellite geodesy with small satellite separation.

%\begin{figure}[t]
%\centering
%\begin{tikzpicture}
%\begin{axis}[ 
%    legend cell align={left},
%    legend style={at={(0.95,0.85)},nodes={scale=0.8,
%        transform shape}}, ylabel style={align=left},
%    ymode=log,xmode=log,
%    ylabel style={align=center}, xlabel={Satellite Separation (km)},
%   % ylabel=1-$\frac{\log \kappa_{\alpha}}{\log \kappa_{1,\mathrm{opt}}}$,
%    ylabel=Necessary $x_T$ \\ improvement 
%  ,domain=0.0005:10
%  ,ymin=10,ymax=10000,
%  xmin=0.0005,xmax=1, 
%width=0.65\columnwidth,legend pos=south east,height=0.65\columnwidth
%]
%  \addplot[mark=.,blue]table[x=dist,y=xc,col sep=comma]{fig_data/sn_req_fig/SQreq_int_OGLPN_50km.csv};
%  % \addplot[mark=.,red]table[x=dist,y=xc,col sep=comma]{fig_data/sn_req_fig/SQreq_OGLPN_imp_50km.csv};
%\end{axis}
%\end{tikzpicture}
%\caption{\textbf{Laser phase noise improvement necessary to reach the quantum noise limit for satellite orbital height of 50 km.} At this orbital height the requirements to be quantum noise limited are not beyond the realms of possibility at small satellite separations. Transitioning to even lower orbital heights would make the requirements less stringent. We again assume laser thermal noise $x_T=1\times10^{-15}$ for the comparison.
%\label{Fig:SNlim_req50k}}
%\end{figure}

\begin{figure}[t]
\includegraphics[width=0.4\textwidth]{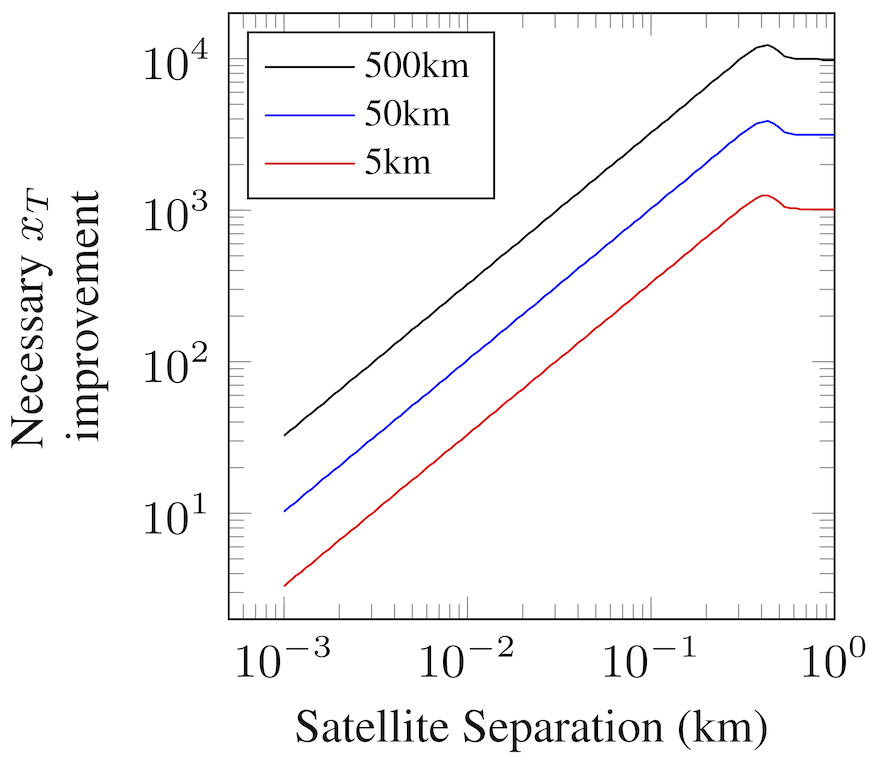}
\caption{\textbf{Laser phase noise improvement necessary to reach the quantum noise limit for different satellite orbital heights.} At orbital heights below 50 km the requirements to be quantum noise limited are not beyond the realms of possibility at small satellite separations. Transitioning to even lower orbital heights makes the requirement even less stringent. We again assume laser thermal noise $x_T=1\times10^{-15}$ for the comparison.
\label{Fig:SNlim_req50k}}
\end{figure}

Indeed if we look at the requirements on the laser phase noise for satellites at lower orbital heights, separated by a few meters, the required laser phase noise improvement to reach the quantum noise limit becomes more attainable as shown in Fig.~\ref{Fig:SNlim_req50k}. The gravitational signal, Eq.~\eqref{eq:arsig}, falls off very rapidly at high frequencies, due to the Bessel function, $K_0(2\pi f h/v_0)$, which becomes small very quickly as $2\pi f h/v_0$ grows. This explains why transitioning to an orbital height of 50 km reduces the requirement on the laser phase noise. Lower orbital heights shift the frequency range of the gravitational signal to higher frequencies. Due to the different frequency dependence of laser phase noise and quantum noise, quantum noise is more significant at higher frequencies. Thus a mission with a sufficiently low orbital height may be quantum noise limited. Such mission parameters are impossible for mapping the Earth's gravitational field, as the drag experienced 50 km above the Earth would be huge. However, for mapping the gravitational field of other astronomical bodies with less atmosphere, such as the Moon, or small planets, like Pluto, this becomes feasible. For example surface pressure on Mercury is approximately 1$\times10^{-14}$ atm and on the Moon surface pressure is effectively negligible, at around 3$\times10^{-15}$ atm. As the drag force at a given height is linearly proportional to the air pressure, the drag force on these smaller bodies will be much less than that on Earth. 

For mapping the gravitational field of smaller astronomical bodies, we can imagine a mission consisting of a single long satellite with two test masses on board flying at a low orbital height. Both test masses could be placed into free-fall inside vacuum within the satellite and the distance between the two masses is measured with a laser interferometer. This is similar to the set-up used on-board LISA pathfinder~\cite{armano2016sub}. In this case the need for an accelerometer is removed as both masses are in free-fall, leaving only laser phase noise and quantum noise. In this regime the laser phase noise improvement necessary before the quantum noise limit is as small as a factor of 10. Such a mission brings satellite geodesy into the realm where squeezing may be useful.

\end{document}